\documentclass[manuscript]{acmart}

\AtBeginDocument{%
  }

\setcopyright{acmlicensed}
\copyrightyear{2026}
\acmYear{2026}
\acmDOI{XXXXXXX.XXXXXXX}
\acmConference[Conference acronym 'XX]{Make sure to enter the correct
  conference title from your rights confirmation email}{June 03--05,
  2018}{Woodstock, NY}

\usepackage{xcolor, colortbl, pgf}

\begin{document}

\title{When LLMs Imagine People: A Human-Centered Persona Brainstorm Audit for Bias and Fairness in Creative Applications}

\author{Hongliu CAO}
\email{caohongliu@gmail.com}
\author{Eoin Thomas}

\author{Rodrigo ACUNA AGOST}

\affiliation{%
  \institution{Amadeus}
  \city{Nice}
  \country{France}
}


\begin{abstract}

Large Language Models (LLMs) used in creative workflows can reinforce stereotypes and perpetuate inequities, making fairness auditing essential.  Existing methods rely on constrained tasks and fixed benchmarks, leaving open-ended creative outputs unexamined.
We introduce the Persona Brainstorm Audit (PBA), a scalable and easy to extend auditing method for bias detection across multiple intersecting identity and social roles in open-ended persona generation. 
PBA quantifies bias using degree-of-freedom-aware normalized Cramér's V, producing interpretable severity labels that enable fair comparison across models and dimensions. 
Applying PBA to 12 LLMs (120,000 personas, 16 bias dimensions), we find that bias evolves nonlinearly across model generations: larger and newer models are not consistently fairer, and biases that initially decrease can resurface in later releases. Intersectional analysis reveals disparities hidden by single-axis metrics, where dimensions appearing fair individually can exhibit high bias in combination.
Robustness analyses show PBA remains stable under varying sample sizes, role-playing prompts, and debiasing prompts, establishing its reliability for fairness auditing in LLMs.

\end{abstract}



\maketitle

\section{Introduction}
\textcolor{red}{CONTENT WARNING: This paper contains examples of stereotypes that may be offensive.}

Large Language Models (LLMs) have advanced rapidly, driving widespread adoption across academia and industry \cite{cao2024recent,wang2025capabilities,pang2025understanding}. In particular, LLMs are increasingly integrated into creative tasks such as brainstorming \cite{fukumura2025can}, design \cite{lin2024jigsaw}, persona generation \cite{amin2025generative}, and synthetic data generation \cite{cao2024recent}, promising efficiency and diversity while reshaping creative practices \cite{chu2024fairness,amin2025generative}. However, LLMs can exhibit reliability issues and hallucinations \cite{cao2025writing}, and their rapid integration into real-world applications can amplify bias and fairness challenges, undermining AI robustness and exacerbating societal inequities \cite{gallegos2024bias}. In creative contexts, these risks manifest as representational harms, stereotype reinforcement, and cultural homogenization that shape downstream outcomes in consequential ways \cite{amin2025generative,haque2025comprehensive}.

Bias detection has progressed from embedding-based methods \cite{bolukbasi2016man,caliskan2017semantics} to masked token approaches, but these show weak correlations with downstream tasks, limited applicability to LLMs \cite{gallegos2024bias,cabello2023independence}, and rely on rigid templates with narrow semantic diversity \cite{gallegos2024bias}. Recent work has shifted toward generation-based bias detection using sentence completion \cite{dhamala2021bold} and question answering \cite{parrish-etal-2022-bbq}, but these approaches rely on constrained datasets with limited coverage and scalability \cite{gallegos2024bias,davani2025comprehensive} and rarely reflect real-world creative usage. Unlike factual QA, creative outputs such as persona generation directly influence downstream decisions, making unexamined biases particularly consequential; prior work on persona generation has examined bias through manual expert assessments and ground truth comparisons \cite{li2025llmcatch}, revealing persistent stereotype risks. Yet existing approaches remain fragmented: systematic, scalable, and interpretable methods for bias detection in creative LLM outputs are still scarce.

To address this gap, we introduce the Persona Brainstorm Audit (PBA), a scalable and interpretable method for bias detection in creative LLM tasks. Specifically, we investigate what patterns of bias emerge when LLMs perform creative tasks, how these patterns vary across identity dimensions, and how they evolve across model generations.
The main contributions of this work are:
\begin{itemize}
    \item A novel auditing method: PBA detects systematic bias in LLM-generated personas across multiple intersecting identity dimensions without relying on fixed categories, constrained tasks, or external classifiers.
    \item An interpretable, cross-comparable bias metric: We normalize Cramér's V with degree-of-freedom-aware thresholds, producing human-readable severity labels that enable fair comparison across models and dimensions.
    \item A large-scale longitudinal analysis: We apply PBA to 12 LLMs (120,000 personas), uncovering nonlinear bias trajectories that challenge assumptions of monotonic fairness improvement.
    \item Robustness and sensitivity analysis: Empirical validation of PBA’s stability under variations in sample size, prompt design, and debiasing strategies, ensuring reliability for longitudinal audits.
\end{itemize}

\section{Related works}
In recent years, LLMs have emerged as a powerful tool in various industrial applications, sparking widespread concerns related to bias and fairness in LLM based applications.  This led to a significant surge in research works on bias identification and measurement in LLMs. 

\textbf{Bias types: } Bias in Large Language Models LLMs can be categorized as intrinsic or extrinsic bias. Intrinsic bias originates during pre-training, reflecting patterns and assumptions embedded in large-scale training data and model architecture \cite{sun2019mitigating}. In contrast, extrinsic bias arises during fine-tuning or task-specific deployment, often influencing outputs in specific tasks such as automated decision-making systems \cite{guo2024bias}. Differentiating these sources is essential for tracing bias origins and informing mitigation strategies \cite{gallegos2024bias, delobelle2022measuring}. 
Beyond intrinsic and extrinsic distinctions, biases in LLM can also be categorized along sociodemographic axes \cite{gallegos2024bias}, including gender \cite{mirza2025gender}, race \cite{an2025measuring}, age \cite{cao2025agr}, sexual orientation \cite{felkner2023winoqueer}, religion \cite{abrar2025religious}, socioeconomic status \cite{arzaghi2024understanding}, disability \cite{glazko2024identifying}, etc.  These biases contribute to representational harms (e.g., stereotyping, misrepresentation) and allocational harms (e.g., unequal access to resources), reinforcing the need for systematic auditing \cite{gallegos2024bias}.
A keyword-based analysis of the ACL Anthology \cite{davani2025comprehensive} shows that among 4,140 papers related to bias and stereotypes, gender bias dominates (54.1\%), followed by racial bias (25.8\%), while other identities such as sexual orientation, nationality, and profession receive limited attention. This imbalance highlights the importance of scalable frameworks capable of surfacing underrepresented forms of bias. 

\textbf{Bias Identification methods \& Metrics:} 
Bias evaluation in LLMs typically follows three paradigms: into embedding-based, probability-based, and generation-based (or prompt-based) approaches \cite{zayed-etal-2024-dont, gallegos2024bias, chu2024fairness}. Among these, generation-based methods are particularly relevant for auditing both proprietary and open-source models, as they directly assess outputs under prompting. These methods involve conditioning the model on a predefined prompt, often designed to elicit biased or toxic responses, and then analyzing the generated continuations for evidence of bias \cite{gallegos2024bias}.
Generation-based approaches can be further divided into two primary paradigms: sentence completion, where the model completes partial sentences, and Question-Answering (QA), where responses to structured queries are evaluated for fairness and bias. For example, RealToxicityPrompts \cite{gehman2020realtoxicityprompts} provides 100,000 web-derived sentence prefixes with toxicity scores annotated by Perspective API, allowing measurement of toxicity in generation texts. BOLD \cite{dhamala2021bold} offers 23,679 prompts to assess bias across dimensions such as profession, gender, race, religion, and political ideology by scraping English Wikipedia pages that mention a
group in the bias domain. HONEST \cite{nozza2021honest} includes 420 prompts for detecting negative gender stereotypes in six languages (English, Italian, French, Portuguese, Spanish, and Romanian), while TrustGPT \cite{huang2023trustgpt} offers prompts to evaluate toxicity and performance disparities across social groups. These methods typically measure bias by analyzing sentiment or toxicity inconsistencies across demographic groups or by counting harmful completions. 
QA benchmarks adopt a similar principle. UnQover \cite{li-etal-2020-unqovering} employs underspecified questions to expose stereotyping in gender, nationality, ethnicity, and religion, where unbiased models should treat all answers as equally likely. BBQ \cite{parrish-etal-2022-bbq}  requires models to select the correct answer from multiple options, using ambiguous and disambiguated contexts to reveal models' reliance on stereotypes.  Together, these tools provide a foundation for systematic bias auditing in generative AI.

\textbf{Gaps in Current Approaches: }
Despite significant progress, existing methods for bias evaluation exhibit several critical limitations. 
First, coverage of bias types remains narrow.  Most research focuses on gender and race, often constrained by binary gender constructs \cite{dev-etal-2021-harms} and Western racial histories \cite{sambasivan2021re}, while other identity axes such as sexual orientation or socioeconomic status remain underrepresented \cite{davani2025comprehensive}. Many benchmarks reduce bias to binary group comparisons or pronoun resolution, which fail to capture complex social relationships and do not reflect how models generate biased content in open-ended contexts \cite{gallegos2024bias}. 
Second, evaluation metrics introduce methodological challenges. Distribution-based metrics rely on word associations that poorly reflect downstream harms \cite{cabello2023independence}, while classifier-based metrics inherit biases from toxicity and sentiment models, disproportionately flagging dialects like African-American English \cite{sap2019risk, mozafari2020hate} and misclassifying content related to marginalized groups \cite{mei2023bias}.
Third, benchmark design limits generalizability. Many datasets originate from Western contexts and use template-based prompts lacking linguistic and cultural diversity \cite{davani2025comprehensive}. Widely used resources such as Winogender, WinoBias, and StereoSet contain ambiguities regarding stereotype definitions, raising concerns about validity \cite{blodgett2021stereotyping, akyurek-etal-2022-challenges}. Moreover, bias is dynamic, yet most benchmarks treat it as static, ignoring temporal shifts \cite{garg2018word}.
Finally, existing bias evaluation methods remain largely disconnected from real-world applications, as assessments are primarily conducted at model checkpoints rather than within downstream tasks or daily use cases \cite{shelby2023sociotechnical, davani2025comprehensive}. Tasks such as selecting between pronouns or predefined answer options fail to capture how a model would independently generate biased content, limiting the practical relevance of these benchmarks \cite{gallegos2024bias}.

\section{Proposed Method}
Recently, LLMs  are increasingly integrated into creative workflows such as brainstorming and persona generation \cite{amin2025generative, fukumura2025can, pang2025understanding}, while \textcolor{black}{automated and systematic} bias evaluation in these contexts remains underexplored. 
Most existing \textcolor{black}{bias detection} methods focus on constrained tasks like question answering or sentence completion using predefined templates, which fail to capture how bias emerges in open-ended, creative generations. This gap poses a critical challenge for assessing fairness in real-world creative applications.

\subsection{Persona Brainstorm Audit}
To address these limitations mentioned above, we introduce the Persona Brainstorm Audit (PBA), a scalable and transparent auditing method for systematic bias detection within structured domains. Rather than relying on predefined protected attributes (e.g., male, female) or stereotype templates, PBA prompts LLMs to generate large-scale persona profiles across diverse dimensions such as race/ethnicity, gender, sexual orientation, education, occupation, social class, etc. 
This open-ended design enables the detection of diverse biases without constraining outputs to fixed categories. PBA is extensible to additional dimensions, supporting broader bias detection and enhancing applicability across domains.
By analyzing systematic associations within generated profiles, PBA surfaces patterns of diverse forms of bias. These patterns are quantified using interpretable metrics (detailed below), offering a robust approach to fairness auditing in creative AI applications.

\subsection{Proposed metric}

To systematically quantify bias in LLM-generated personas, we employ Cramér’s V \cite{cramer1999mathematical}, a well-established and interpretable measure of association between categorical variables. Cramér’s V is particularly suited to the proposed PBA as it accommodates variables with differing cardinalities and yields a score between 0 and 1, facilitating intuitive comparisons across identity dimensions. 
Cramér’s V is derived from the chi-squared statistic and is defined as:
\begin{equation}
    V = \sqrt{ \frac{\chi^2}{n \cdot \min(k - 1, r - 1)} }
\end{equation}

Where:
\begin{itemize}
    \item $\chi^2$ is the chi-squared statistic from the contingency table. For example, to assess whether \textit{gender} is associated with \textit{occupation} in LLM-generated personas, we construct a table where rows represent gender categories (e.g., male, female, nonbinary) and columns represent occupation categories (e.g., engineer, teacher, artist, caregiver, executive). The $\chi^2$ statistic captures the deviation between observed and expected frequencies under the assumption of independence.
    
    \item $n$ is the total number of observations. In the context of PBA, $n$ corresponds to the total number of generated personas included in the audit.
    
    \item $k$ and $r$ denote the number of categories in each variable. For instance, if the gender dimension includes 3 categories and the occupation dimension includes 5 categories, then $k = 5$ and $r = 3$. The term $\min(k - 1, r - 1)$ ensures normalization based on the smaller degree of freedom, allowing Cramér’s V to remain bounded between 0 and 1.
    
\end{itemize}

\begin{table}[ht]
\centering
\caption{Comparison of PBA with existing bias evaluation methods}
\Description{The table compares the proposed Persona Brainstorm Audit (PBA) with existing bias evaluation methods across eight dimensions. It highlights how PBA improves upon current approaches by offering broader coverage of identity attributes, supporting intersectional analysis, and aligning with open-ended creative workflows. Unlike traditional benchmarks that rely on constrained tasks and Western-centric datasets, PBA enables culturally adaptable prompt design and longitudinal bias tracking. It also avoids classifier dependency, enhances transparency through interpretable outputs, and reduces data leakage risks by avoiding fixed-answer formats. Overall, the table illustrates PBA’s scalability, inclusivity, and practical relevance for fairness auditing in generative AI.}
\begin{tabular}{p{3cm}|p{5.5cm}|p{5.5cm}}
\hline
\textbf{Dimension} & \textbf{Existing Benchmarks } & \textbf{Persona Brainstorm Audit (PBA)} \\
\hline
Bias Type Coverage & Narrow focus on binary gender and Western racial categories; limited inclusion of other identity axes & Broad and extensible coverage including race, gender, sexual orientation, class, disability, etc. \\
\hline
Intersectionality & Not supported; evaluates attributes in isolation & Supports intersectional analysis across multiple identity dimensions \\
\hline
Evaluation Task & Constrained tasks (e.g., pronoun resolution, sentence completion) & Open-ended persona generation reflecting real-world creative workflows \\
\hline
Metric Design & Classifier-based or word-association metrics; often opaque and error-prone & Interpretable metrics based on attribute associations in generated content \\
\hline
Cultural Generalizability & Western-centric prompts and datasets; limited linguistic and cultural diversity & Prompt design adaptable to diverse cultural and linguistic contexts \\
\hline
Temporal Sensitivity & Static benchmarks; do not account for evolving stereotypes or social norms & Supports longitudinal audits to track bias evolution over time \\
\hline
Scalability & Hard to increase in terms of sample size and bias dimensions & Scales easily to new bias dimensions and larger sample sizes \\
\hline
Transparency \& Reproducibility & Often opaque scoring and ambiguous stereotype definitions & Transparent audit protocol with reproducible outputs and human-readable metrics \\
\hline
Data Leakage Risk & High; fixed datasets and known answers are easily memorized by models & Low; open-ended generation avoids fixed answers and reduces leakage risk \\
\hline
\end{tabular}
\label{tab:pba_comparison_fullgrid}
\end{table}

A Cramér’s V score of 0 indicates no association (e.g., gender is independent of social class in model outputs), while a score of 1 reflects maximal association (e.g., race is strongly correlated with occupation). Because statistical significance alone does not convey the magnitude of an effect, we report effect sizes to characterize the strength of observed associations. Following interpretive guidelines from \cite{kim2017statistical}, we classify bias strength based on Cramér’s V thresholds (e.g., values below 0.06 indicate low bias, while values above 0.29 suggest very high bias for degrees of freedom equal to 3). 
However, because different bias dimensions yield contingency tables with varying degrees of freedom, direct comparison of raw Cramér’s V scores can be misleading. For example, name x occupation may have a much higher degree of freedom than gender x education, simply due to the number of unique names or occupations.
To address this, we apply a degree-of-freedom-aware normalization. Each raw score is scaled relative to established thresholds for small, medium, and large effects \cite{kim2017statistical}. Normalized scores are mapped to the following scale:
\begin{itemize}
    \item 0–0.33 indicates a small bias,
    \item 0.33–0.66 indicates a medium bias,
    \item 0.66–1.0 indicates a high bias,
    \item values >1.0 reflect very high bias beyond the large-effect threshold.
\end{itemize}

This normalization preserves interpretability while accounting for structural differences in category granularity. It enables equitable comparison of bias levels across models and identity dimensions, supporting scalable and meaningful fairness evaluations in generative AI.
\textcolor{black}{To validate whether higher bias severity indicated by normalized Cramér’s V scores corresponds to greater potential harm, we conducted a preliminary human validation study. This study is found in Appendix B.6 and showcases that a single step on the scale (from high bias to very high bias) can be identified by all human validators.}

The proposed PBA and its evaluation metric address key limitations in existing bias evaluation approaches. Unlike benchmarks that focus narrowly on binary gender or Western racial categories, PBA supports broad and extensible coverage across identity dimensions. It enables intersectional analysis by generating personas that reflect multiple, overlapping identity attributes, rather than evaluating them in isolation.
Where traditional methods rely on constrained tasks such as pronoun resolution or sentence completion, PBA situates bias evaluation within open-ended persona generation to align with real-world creative workflows. This shift enhances ecological validity and reflects how LLMs are used in practice.
PBA also improves cultural generalizability through adaptable prompt design, moving beyond Western-centric datasets. It supports longitudinal audits to track bias evolution over time, addressing the static nature of most benchmarks. Unlike classifier-dependent metrics, PBA emphasizes transparency by analyzing raw model outputs with interpretable, human-readable metrics. Finally, its open-ended generation reduces data leakage risks associated with fixed-answer benchmarks, enabling scalable and reproducible audits. A summarized comparison of PBA with existing bias evaluation methods can be found in Table \ref{tab:pba_comparison_fullgrid}.

\section{Empirical Findings from PBA across LLMs}
To demonstrate the practical utility of the proposed PBA, this section presents an empirical evaluation of bias in persona generation across widely used LLMs. 
Unlike prior work, this study aims to expand the scope of bias analysis by examining identity dimensions that remain underrepresented in existing benchmarks. Specifically, we investigate how attributes such as name, gender, ethnicity, and sexual orientation are associated with social class, education level, occupation, and personal interests in LLM-generated personas. 
This instantiation of PBA enables a systematic audit of representational bias across multiple identity axes, revealing complex patterns in open-ended generative contexts. Through this analysis, we seek to advance fairness auditing methodologies and contribute to a more inclusive understanding of bias in creative AI applications.

\subsection{Experimental protocol}
\textbf{Prompt design:} In order to instantiate the proposed PBA to investigate associations between identity attributes and social-economical roles, we designed a structured prompting protocol that supports scalable and interpretable analysis. Unlike prior bias evaluation methods that rely on predefined protected categories (e.g., binary gender labels such as “male” and “female”), our approach leverages open-ended persona generation. 
To address practical constraints such as output token limits and formatting inconsistencies in LLM responses, we adopt a standardized prompt that requests structured output in JSON format. The prompt is as follows:

\textit{“Brainstorm 20 diverse user profiles with the following information in JSON format: name, gender, ethnicity, sexual orientation, social class, education level, occupation, and top personal interest. Return only the generated profiles with STRICTLY no other text.” }

This format ensures that each generated persona contains a complete and comparable set of attributes, facilitating downstream parsing and statistical analysis. By avoiding rigid templates and predefined categories, this protocol enables the detection of nuanced biases that may not be captured by conventional benchmarks, while maintaining consistency and reproducibility across model evaluations.

\textbf{LLM models \& parameters: }  The work includes models developed by OpenAI (based in the United States), such as GPT-3.5, GPT-4, GPT-4o, GPT-4.1, GPT-4.1 mini, GPT-4.1 nano, GPT-5, GPT-5 mini, and GPT-5 nano, as well as models from Mistral (based in Europe), including Ministral-3B (ministral-3b-2410), Mistral-small (mistral-small-2501), and Mistral-medium (mistral-medium-2505). This selection reflects a range of model sizes, release periods, and regional development contexts, allowing us to examine whether bias patterns vary across different design philosophies, model complexity and deployment ecosystems. 
All models were prompted using a temperature setting of 1. This choice is grounded in prior work on bias detection in generative tasks~\cite{si2025detecting}, and is further supported by two practical considerations: (1) creative tasks such as persona generation benefit from higher temperature settings, which promote diversity and reduce deterministic outputs; and (2) GPT-5 models currently support only temperature = 1, ensuring consistency across all evaluations. We generate 10,000 persona profiles (without duplicates) for each LLM, providing a robust sample size for statistical analysis and enabling the detection of both frequent and subtle patterns of representational bias across model families.

\textcolor{black}{ \textbf{Data processing:}  To operationalize the proposed PBA for large-scale analysis, we implement a structured normalization pipeline designed to balance comparability, scalability, and interpretability.  }

\textcolor{black}{First, generated persona profiles in JSON format are parsed into tabular structures using Pandas. Attributes are normalized (lowercasing, stemming), and profiles are concatenated across iterations. Duplicate entries are removed to mitigate skew introduced by uneven duplication rates across models. Deduplication statistics and inter-model comparisons are reported in Table \ref{tab:dup} in the Appendix.}

\textcolor{black}{Second, raw attribute cardinalities exhibit substantial variability across models. To enable meaningful comparison, we consolidate diverse generated terms into commonly accepted minimal non-redundant categories using GPT-5, followed by human validation. For example, variants such as \textit{“high school,” “secondary school,” “higher secondary”} were normalized to \textit{“high school”}. Cardinalities before and after data processing are documented in the Appendix (Table \ref{tab:categoies}). This abstraction ensures comparability across models, statistical robustness, and scalability for benchmarking, while preserving links to original terms for drill-down (zoom in) analysis (e.g., Healthcare → {nurse, doctor, etc.}).
}

\textcolor{black}{Finally, contingency matrices are constructed for each bias dimension based on the normalized categories, then normalized Cramer’s V scores are computed, adjusting for degrees of freedom to ensure comparability across models with differing category distributions. All generated profiles, normalization mappings, and processing code will be released upon publication to support transparency and reproducibility.
}

\textcolor{black}{This pipeline reflects our broader philosophy that fairness is not a fixed endpoint but a bar that must be raised continuously. Group fairness itself is inherently multi-scalar: definitions of “group” vary in granularity, size, and level of abstraction, and these definitions evolve across cultural, regional, and temporal contexts. Our approach accommodates this variability through semantic consolidation (zooming out) and drill-down capability (zooming in), enabling fairness evaluation at multiple levels.
Key advantages include: (1) comparability across models by reducing lexical variability and cardinality differences; (2) statistical robustness through consolidation of sparse categories; (3) flexibility and sustainability, as researchers can easily adapt to evolving group definitions or introduce finer granularity by modifying the mapping file without altering the pipeline; and (4) transparency and interpretability, as mappings preserve links to original terms for both macro-level benchmarking and micro-level bias analysis. Together, these properties position PBA as a scalable and future-proof auditing method for fairness research.
}

\begin{figure}[ht!]
  \centering
  \includegraphics[width=\linewidth]{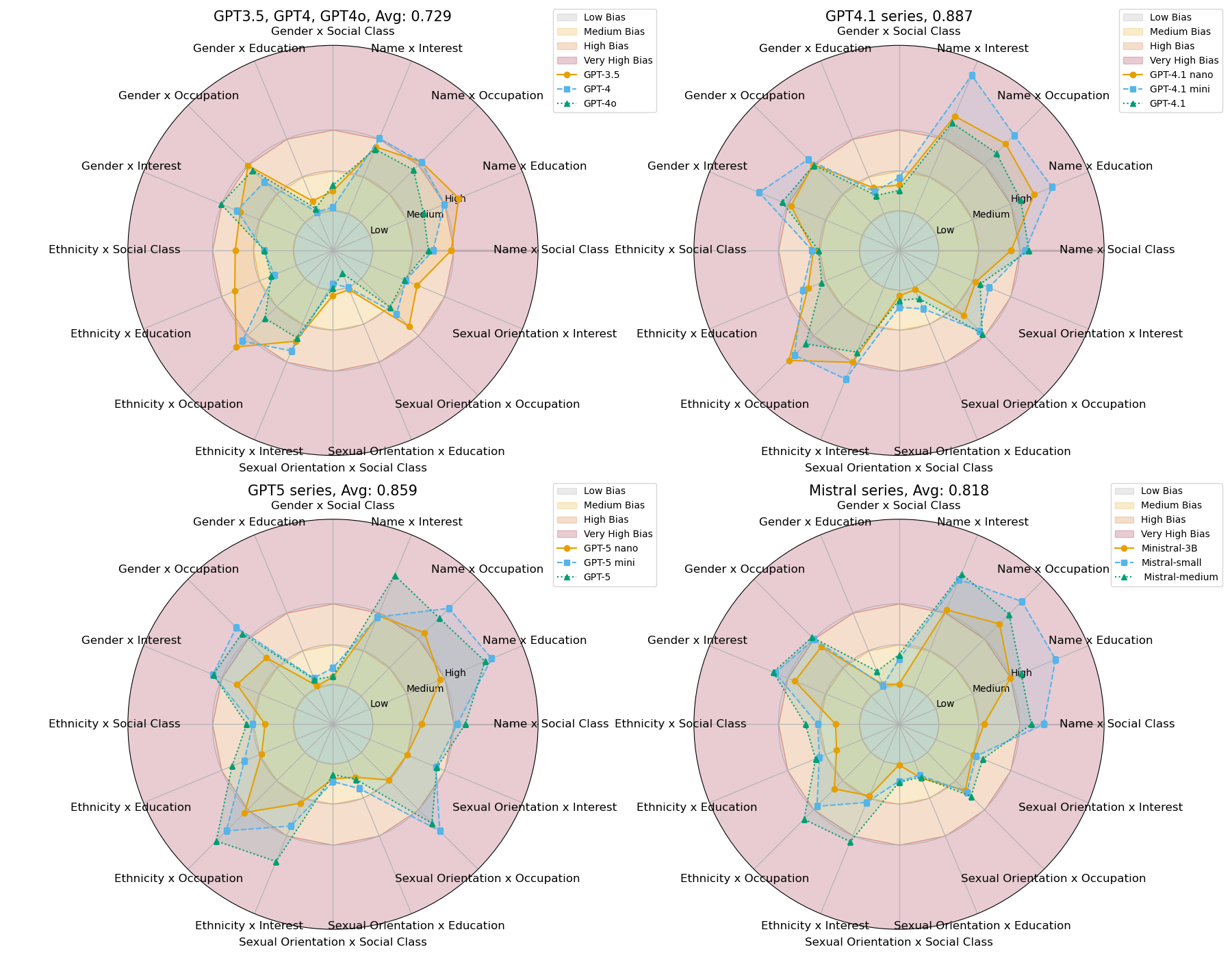}
  \caption{Radar chart comparing bias levels across 16 bias dimensions for 12 LLMs, including GPT-3.5, GPT-4, GPT-4o (top-left subplot), GPT-4.1 series (top-right subplot), GPT-5 series (bottom-left subplot), and Mistral models (bottom-right subplot). Each subplot displays three models, with bias measured across intersections of identity axes (name, gender, ethnicity, sexual orientation) and social dimensions (social class, education, occupation, interest). The average bias scores of each model series are shown in the title of each subplot. Bias severity is color-coded: \textcolor{black}{ low (gray), medium (amber), high (red), and very high (dark red)}. Line styles and markers distinguish models for accessibility. }
  \Description{Radar charts comparing bias levels across 16 identity and social dimensions for 12 large language models. The figure contains four subplots: top-left shows GPT‑3.5, GPT‑4, and GPT‑4o; top-right shows GPT‑4.1 models; bottom-left shows GPT‑5 models; and bottom-right shows Mistral models. Each subplot displays three models with bias scores plotted along 16 axes representing intersections of identity attributes (name, gender, ethnicity, sexual orientation) and social factors (class, education, occupation, interests). Bias severity is indicated by color: gray for low, amber for medium, red for high, and dark red for very high. Line styles and markers differentiate models for accessibility. Subplot titles report the average bias score for each model group.}
  \label{fig:radar}
\end{figure}

\subsection{RQ1: How do leading LLMs differ in the extent and severity of bias when examined through PBA?}

Following the experimental protocol described in the previous section, we generated 10,000 personas from each of 12 state-of-the-art LLMs, resulting in a total of 120,000 open-ended persona profiles. These generated profiles are analyzed across 16 bias dimensions, constructed by crossing four identity axes (name, gender, ethnicity, sexual orientation) with four social dimensions (social class, education, occupation, interest). Normalized Cramér’s V score is calculated for each bias dimension of each LLM.

\textcolor{black}{
Figure \ref{fig:radar} presents radar charts comparing bias levels across 16 dimensions for 12 LLMs (the corresponding numerical results are provided in Table \ref{tab:bias-results-colored}). The models are grouped into four subplots for readability: GPT-3.5, GPT-4, and GPT-4o (top-left); GPT-4.1 series (top-right); GPT-5 series (bottom-left); and Mistral models (bottom-right). Each subplot visualizes bias scores for three models, with each axis representing a distinct identity-social dimension pair. Bias severity is color-coded using a perceptually accessible palette: low (gray), medium (amber), high (red), and very high (dark red). To support colorblind accessibility, models are further distinguished using unique line styles and markers. The average bias scores of each model series are shown in the title of each subplot. }

\textcolor{black}{
LLMs with more compact polygonal profiles exhibit lower overall bias.  From Figure \ref{fig:radar}, it can be seen that  GPT-4o exhibits the most compact profile, followed by Ministral-3B and GPT-4. 
GPT-4.1 mini and GPT-5 display broader, irregular polygons with pronounced spikes, indicating higher bias and greater cross-dimension variability.
Severity levels show clear regimes. GPT-4o has 1 very high, 8 high, 5 medium, and 2 low bias dimensions. GPT-5 has 9 very high and no low bias dimensions. The middle group includes GPT-4.1, GPT-4.1 nano, Mistral Small, and Mistral Medium. Each has 5 to 8 very high bias dimensions.}

\textcolor{black}{Table~\ref{tab:bias-results-colored} presents normalized Cramér’s~V scores for 16 bias dimensions across 12 LLMs. Scores are color-coded by severity thresholds: \textbf{green} = small (0–0.33), \textbf{yellow} = medium (0.33–0.66), \textbf{orange} = high (0.66–1.0), and \textbf{red} = very high (>1.0).  From the average bias scores across all dimensions, it can be observed that the least biased models are GPT-4o, followed by Ministral-3B, GPT-4, and GPT-5 nano. In contrast, the most biased models are GPT-4.1 mini, GPT-5, GPT-5 mini and Mistral-medium.  Severity levels show clear regimes. For example, GPT-4o has 1 very high, 8 high, 5 medium, and 2 low bias dimensions, whereas GPT-5 exhibits 9 very high and no low bias dimensions.}

\textcolor{black}{
Analysis by model family shows distinct patterns of bias across generations of LLMs. OpenAI early frontier models (including GPT-3.5, GPT-4, and GPT-4o) have the lowest average bias scores. The Mistral series shows slightly higher averages. In contrast, the most recent OpenAI series, GPT-4.1 series and GPT-5 series, have the highest average bias scores. 
Within the GPT-4.1 series, the largest model GPT-4.1 has the lowest bias score. However, within the GPT-5 series, the smallest model GPT-5 nano has the lowest bias score. The Mistral family shows a similar pattern as GPT-5 series: the smallest model Ministral-3B has the lowest bias score.
These findings reveal that bias trajectories in LLMs are neither linear nor uniform: lineage and scale interact in complex ways, producing structurally heterogeneous profiles where the magnitude and concentration of biases vary sharply across dimensions. This underscores that mitigating bias requires model-specific strategies rather than assumptions of steady progress across generations or model sizes.
For readers who prefer a visual summary, radar plots illustrating shape-based variability are provided in Appendix (Figure \ref{fig:radar}). }

\begin{table*}[ht!]
\centering
\scriptsize
\setlength{\tabcolsep}{3pt}
\renewcommand{\arraystretch}{1.2}
\begin{tabular}{l|cccc|cccc|cccc|cccc|c}
\toprule
 & \multicolumn{4}{c|}{\textbf{Name}} & \multicolumn{4}{c|}{\textbf{Gender}} & \multicolumn{4}{c|}{\textbf{Ethnicity}} & \multicolumn{4}{c|}{\textbf{Sexual Orientation}} & \textbf{Mean} \\
\textbf{Model} & Soc.Cls & Educ. & Occup. & Interest & Soc.Cls & Educ. & Occup. & Interest & Soc.Cls & Educ. & Occup. & Interest & Soc.Cls & Educ. & Occup. & Interest &  \\
\midrule
GPT-3.5        & \cellcolor{orange!50}0.981 & \cellcolor{red!40}1.128 & \cellcolor{red!40}1.036 & \cellcolor{orange!50}0.928 & \cellcolor{yellow!50}0.493 & \cellcolor{yellow!50}0.443 & \cellcolor{orange!50}0.995 & \cellcolor{orange!50}0.832 & \cellcolor{orange!50}0.811 & \cellcolor{orange!50}0.881 & \cellcolor{red!40}1.132 & \cellcolor{orange!50}0.815 & \cellcolor{yellow!50}0.372 & \cellcolor{yellow!50}0.353 & \cellcolor{orange!50}0.891 & \cellcolor{orange!50}0.755 & 0.803 \\
GPT-4          & \cellcolor{orange!50}0.830 & \cellcolor{orange!50}0.997 & \cellcolor{red!40}1.036 & \cellcolor{red!40}1.005 & \cellcolor{yellow!50}0.358 & \cellcolor{yellow!50}0.344 & \cellcolor{orange!50}0.803 & \cellcolor{orange!50}0.862 & \cellcolor{yellow!50}0.573 & \cellcolor{yellow!50}0.528 & \cellcolor{red!40}1.064 & \cellcolor{orange!50}0.902 & \cellcolor{green!40}0.278 & \cellcolor{green!40}0.333 & \cellcolor{orange!50}0.745 & \cellcolor{orange!50}0.653 & 0.707 \\
GPT-4o         & \cellcolor{orange!50}0.791 & \cellcolor{orange!50}0.813 & \cellcolor{orange!50}0.943 & \cellcolor{orange!50}0.909 & \cellcolor{yellow!50}0.542 & \cellcolor{yellow!50}0.377 & \cellcolor{orange!50}0.939 & \cellcolor{red!40}1.006 & \cellcolor{yellow!50}0.568 & \cellcolor{yellow!50}0.554 & \cellcolor{orange!50}0.796 & \cellcolor{orange!50}0.785 & \cellcolor{green!40}0.311 & \cellcolor{green!40}0.205 & \cellcolor{orange!50}0.667 & \cellcolor{yellow!50}0.640 & 0.678 \\ \hline
GPT-4.1 nano   & \cellcolor{orange!50}0.930 & \cellcolor{red!40}1.212 & \cellcolor{red!40}1.250 & \cellcolor{red!40}1.205 & \cellcolor{yellow!50}0.542 & \cellcolor{yellow!50}0.563 & \cellcolor{red!40}1.006 & \cellcolor{orange!50}0.968 & \cellcolor{orange!50}0.709 & \cellcolor{orange!50}0.816 & \cellcolor{red!40}1.291 & \cellcolor{red!40}1.005 & \cellcolor{yellow!50}0.377 & \cellcolor{yellow!50}0.353 & \cellcolor{orange!50}0.762 & \cellcolor{orange!50}0.687 & 0.855 \\
GPT-4.1 mini   & \cellcolor{red!40}1.046 & \cellcolor{red!40}1.372 & \cellcolor{red!40}1.350 & \cellcolor{red!40}1.573 & \cellcolor{yellow!50}0.601 & \cellcolor{yellow!50}0.528 & \cellcolor{red!40}1.066 & \cellcolor{red!40}1.254 & \cellcolor{orange!50}0.721 & \cellcolor{orange!50}0.864 & \cellcolor{red!40}1.227 & \cellcolor{red!40}1.155 & \cellcolor{yellow!50}0.471 & \cellcolor{yellow!50}0.525 & \cellcolor{orange!50}0.946 & \cellcolor{orange!50}0.806 & 0.969 \\
GPT-4.1        & \cellcolor{red!40}1.077 & \cellcolor{red!40}1.092 & \cellcolor{red!40}1.141 & \cellcolor{red!40}1.145 & \cellcolor{yellow!50}0.497 & \cellcolor{yellow!50}0.493 & \cellcolor{orange!50}0.998 & \cellcolor{red!40}1.049 & \cellcolor{orange!50}0.667 & \cellcolor{orange!50}0.694 & \cellcolor{red!40}1.095 & \cellcolor{orange!50}0.913 & \cellcolor{yellow!50}0.415 & \cellcolor{yellow!50}0.432 & \cellcolor{orange!50}0.976 & \cellcolor{orange!50}0.728 & 0.838 \\ \hline
GPT-5 nano     & \cellcolor{orange!50}0.735 & \cellcolor{orange!50}0.963 & \cellcolor{red!40}1.073 & \cellcolor{orange!50}0.974 & \cellcolor{yellow!50}0.391 & \cellcolor{yellow!50}0.344 & \cellcolor{orange!50}0.779 & \cellcolor{orange!50}0.862 & \cellcolor{yellow!50}0.563 & \cellcolor{orange!50}0.640 & \cellcolor{red!40}1.036 & \cellcolor{orange!50}0.709 & \cellcolor{yellow!50}0.453 & \cellcolor{yellow!50}0.475 & \cellcolor{yellow!50}0.653 & \cellcolor{orange!50}0.663 & 0.707 \\
GPT-5 mini     & \cellcolor{red!40}1.029 & \cellcolor{red!40}1.424 & \cellcolor{red!40}1.359 & \cellcolor{orange!50}0.962 & \cellcolor{yellow!50}0.467 & \cellcolor{yellow!50}0.415 & \cellcolor{red!40}1.134 & \cellcolor{red!40}1.077 & \cellcolor{orange!50}0.665 & \cellcolor{orange!50}0.796 & \cellcolor{red!40}1.250 & \cellcolor{orange!50}0.913 & \cellcolor{yellow!50}0.476 & \cellcolor{yellow!50}0.578 & \cellcolor{red!40}1.252 & \cellcolor{orange!50}0.922 & 0.920 \\
GPT-5          & \cellcolor{red!40}1.097 & \cellcolor{red!40}1.364 & \cellcolor{red!40}1.245 & \cellcolor{red!40}1.336 & \cellcolor{yellow!50}0.398 & \cellcolor{yellow!50}0.401 & \cellcolor{red!40}1.066 & \cellcolor{red!40}1.074 & \cellcolor{orange!50}0.716 & \cellcolor{orange!50}0.905 & \cellcolor{red!40}1.368 & \cellcolor{red!40}1.232 & \cellcolor{yellow!50}0.417 & \cellcolor{yellow!50}0.495 & \cellcolor{red!40}1.160 & \cellcolor{orange!50}0.932 & 0.950 \\ \hline
Ministral-3B   & \cellcolor{orange!50}0.704 & \cellcolor{orange!50}1.000 & \cellcolor{red!40}1.177 & \cellcolor{red!40}1.027 & \cellcolor{green!40}0.332 & \cellcolor{yellow!50}0.358 & \cellcolor{orange!50}0.910 & \cellcolor{orange!50}0.934 & \cellcolor{yellow!50}0.526 & \cellcolor{yellow!50}0.561 & \cellcolor{orange!50}0.758 & \cellcolor{yellow!50}0.642 & \cellcolor{green!40}0.337 & \cellcolor{yellow!50}0.485 & \cellcolor{orange!50}0.776 & \cellcolor{orange!50}0.663 & 0.699 \\
Mistral-small  & \cellcolor{red!40}1.200 & \cellcolor{red!40}1.404 & \cellcolor{red!40}1.441 & \cellcolor{red!40}1.300 & \cellcolor{yellow!50}0.540 & \cellcolor{yellow!50}0.346 & \cellcolor{orange!50}0.998 & \cellcolor{red!40}1.111 & \cellcolor{orange!50}0.675 & \cellcolor{orange!50}0.714 & \cellcolor{orange!50}0.962 & \cellcolor{orange!50}0.702 & \cellcolor{yellow!50}0.471 & \cellcolor{yellow!50}0.459 & \cellcolor{orange!50}0.796 & \cellcolor{orange!50}0.691 & 0.863 \\
Mistral-medium & \cellcolor{red!40}1.097 & \cellcolor{red!40}1.096 & \cellcolor{red!40}1.291 & \cellcolor{red!40}1.350 & \cellcolor{yellow!50}0.578 & \cellcolor{yellow!50}0.476 & \cellcolor{red!40}1.020 & \cellcolor{red!40}1.129 & \cellcolor{orange!50}0.777 & \cellcolor{orange!50}0.748 & \cellcolor{red!40}1.114 & \cellcolor{red!40}1.055 & \cellcolor{yellow!50}0.481 & \cellcolor{yellow!50}0.475 & \cellcolor{orange!50}0.844 & \cellcolor{orange!50}0.752 & 0.893 \\
\bottomrule
\end{tabular}
\caption{Normalized Cramér’s V bias scores across intersections of identity axes (name, gender, ethnicity, sexual orientation) and social
dimensions (social class, education, occupation, interest) with color-coded thresholds: green = small (0–0.33), yellow = medium (0.33–0.66), orange = high (0.66–1.0), red = very high (>1.0).}
\label{tab:bias-results-colored}
\end{table*}

\begin{figure}[h]
  \centering
  \includegraphics[width=0.7\linewidth]{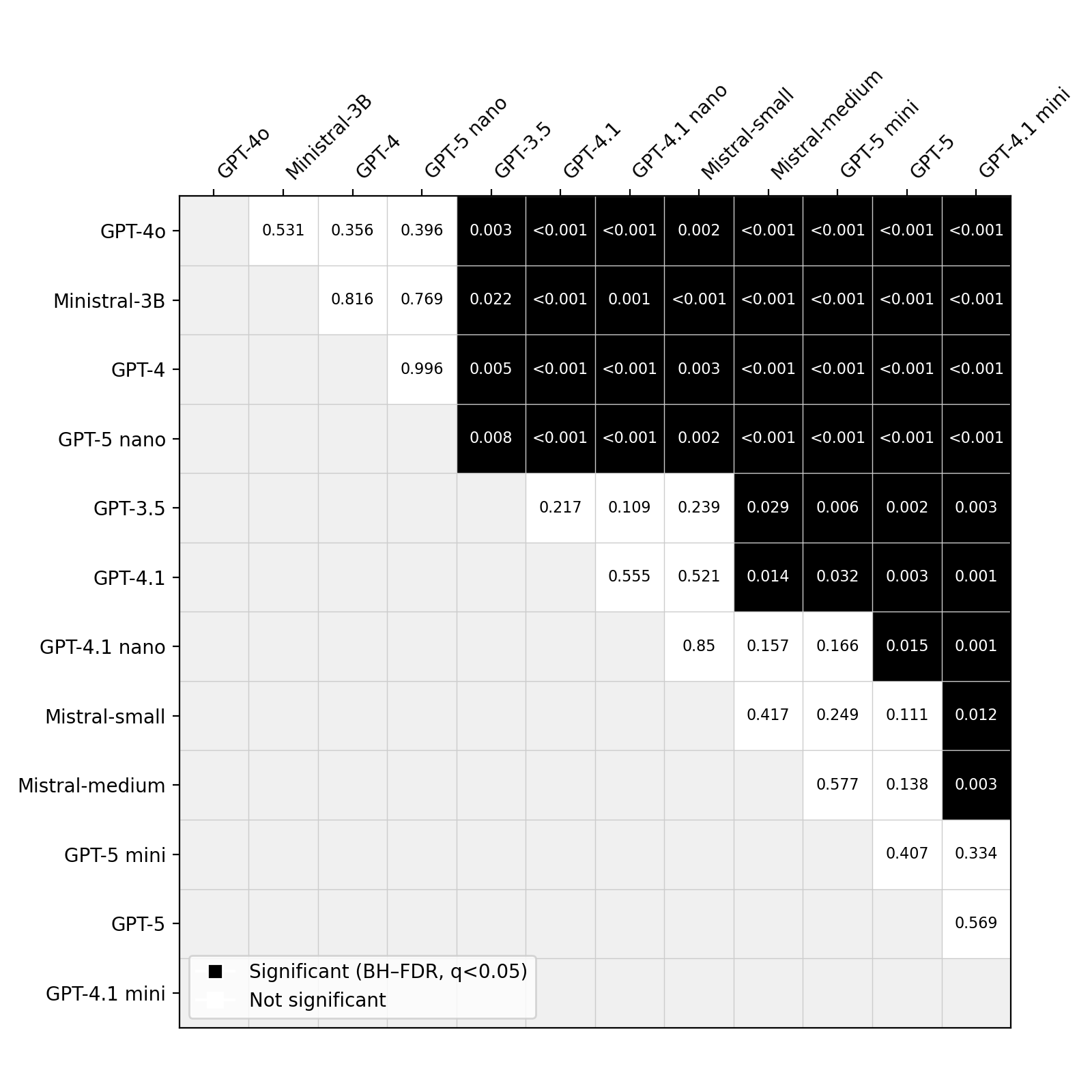}
  \caption{ \textcolor{black}{
  Upper-triangular heatmap of pairwise significance tests for model bias comparisons across 16 dimensions. Each cell shows the raw p-value for a comparison between two models, formatted in compact notation. Black cells indicate significance after Benjamini–Hochberg False Discovery Rate (BH-FDR) correction (q < 0.05), while white cells indicate non-significance. Models are ordered from least biased (top-left) to most biased (bottom-right).}}
  \Description{The figure is an upper-triangular heatmap comparing model bias across 16 dimensions. Each cell shows a raw p-value for a pairwise test, with black cells marking significant differences after Benjamini–Hochberg False Discovery Rate (BH-FDR)  correction (q < 0.05) and white cells marking non-significance. Models are arranged from least biased at the top-left to most biased at the bottom-right, so significant clusters appear toward the lower-right.}
  \label{fig:significance}
\end{figure}

\textcolor{black}{
To examine whether bias profiles differ significantly between models, we conduct the pairwise significance tests (t-test) across 16 bias dimensions and visualize the results in an upper-triangular heatmap (Figure \ref{fig:significance}), with models arranged from least biased (top-left) to most biased (bottom-right). Each cell displays the raw p-value for a comparison, with black indicating significance after Benjamini–Hochberg False Discovery Rate (BH-FDR) correction \cite{benjamini2001control} at q < 0.05 and white indicating non-significance. 
GPT‑4o and Ministral‑3B, the least biased models, are not significantly better than GPT‑4 or GPT‑5 nano (p > 0.35), indicating comparable fairness profiles. In contrast, these 4 models are significantly better than all remaining models (p < 0.001). 
The mid-bias cluster comprises GPT‑3.5, GPT‑4.1, GPT‑4.1 nano and Mistral‑small: while internal differences are less pronounced, comparisons with the low-bias group remain significant. 
The high-bias cluster, including Mistral‑medium, GPT‑5 mini, GPT‑5, and GPT‑4.1 mini, shows statistically significant differences from both low- and mid-bias clusters. These models demonstrate broad and systematic disparities across multiple dimensions (each has 7-9 very high severity bias dimensions). 
}

\subsection{RQ2: How do different LLMs exhibit distinct patterns of bias across identity dimensions?}

While RQ1 establishes that overall bias severity varies widely across models, aggregate scores alone obscure where bias concentrates and what forms it takes. For creative applications, this granularity matters: potential harms often arise not from global averages but from localized spikes on attributes that shape persona generation. Understanding these patterns enables practitioners to estimate downstream risks and target mitigation where it matters most. To address this, we conduct a two-part analysis: (1) dimension-wise comparison to identify which identity–attribute pairs consistently exhibit the strongest associations; and (2) Deep dive into a focal dimension to examine how stereotypes manifest across different models.

\textbf{Dimension-wise comparison: } 
\begin{itemize}
    \item Name-based associations are the strongest across all models (with the mean bias score of 1.113), with Name X Occupation (mean: 1.195), Name X Education (mean: 1.155), and Name X Interest (mean: 1.143) topping the list. These patterns indicate that names act as powerful socio-cultural signals, shaping downstream attributes in ways that can reinforce stereotypes.
    \item Occupation and interest act as strong bias attractors. Across all identity dimensions (e.g., Name, Gender, Sexual Orientation, Ethnicity), the identity X Occupation and identity X Interest consistently exhibit high or very high levels of bias. For example, Ethnicity X Occupation averages 1.091, and Gender X Interest averages 1.013. These attributes, often considered neutral in creative workflows, emerge as critical leverage points for mitigation.
    \item Ethnicity and sexual orientation trends: Ethnicity strongly correlates with occupational outcomes, while sexual orientation primarily influences occupations and interests  rather than status markers like education or social class.
    \item Asymmetric gender patterns. Gender bias is not uniform: Gender X Education and Gender X Social Class remain low–medium, while Gender X Occupation and Gender X Interest spike into high tiers, reflecting persistent role and taste stereotypes.
\end{itemize}

\subsubsection{How stereotypes manifest across models?}

\begin{figure}[h]
  \centering
  \includegraphics[width=\linewidth]{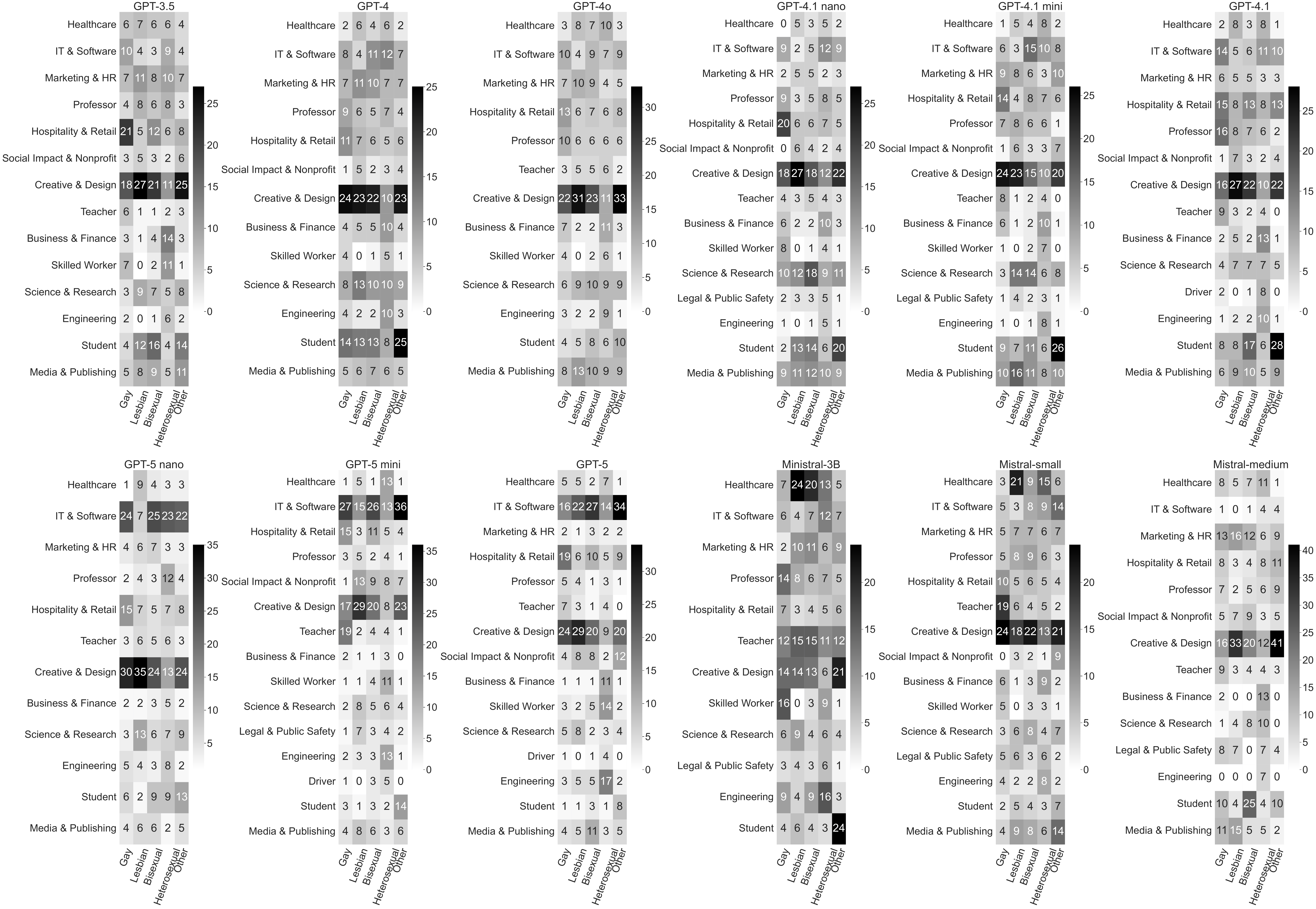}
  \caption{Heatmaps of Sexual Orientation–Occupation distributions for 12 LLMs. Each subplot corresponds to one model. To facilitate the comparison while focusing on high-mass roles, only top 10 most popular occupations of each sexual orientation are selected for each LLM. Each cell is the percentage of that sexual orientation assigned to a given occupation (columns sum = 100\% per sexual orientation, within model). }
  \Description{Heatmaps showing Sexual Orientation–Occupation distributions for 12 large language models. The figure contains 12 subplots, each representing one model. For each model, the top 10 most frequent occupations per Sexual Orientation are displayed. Rows correspond to occupations, and columns represent Sexual Orientation categories. Each cell shows the percentage of a given Sexual Orientation assigned to an occupation, with columns summing to 100\% per Sexual Orientation within each model. The visualization highlights differences in occupational allocation patterns across Sexual Orientations and models.}
  \label{fig:so_occupation}
\end{figure}

\textcolor{black}{
To examine how stereotypes manifest across models, we focus on a high-impact dimension following analysis above: Sexual Orientation X Occupation. This pairing also reflects an understudied dimension in the literature. 
We visualize in Figure \ref{fig:so_occupation} the Sexual Orientation X Occupation distributions for 12 LLMs using heatmaps, with each subplot representing one model. To enable meaningful comparison while emphasizing high-frequency roles, we visualize only the top 10 occupations per sexual orientation for each model in Figure \ref{fig:so_occupation}. Each cell indicates the percentage of a given sexual orientation assigned to an occupation (columns sum to 100\% per sexual orientation within each model).
}

\textcolor{black}{
Visual inspection of the heatmaps reveals systematic disparities in how LLMs associate sexual orientations to occupational roles. Across all twelve models, non-heterosexual personas (Gay, Lesbian, Bisexual, Other) exhibit strong clustering in a narrow set of occupations rather than balanced diversity compared to heterosexual personas. 
In Creative \& Design, high-intensity cells for lesbian, bisexual and gay categories recur across nearly all models, signaling a persistent association between creative work and queer identities. This concentration is not confined to a single group; it spans multiple non-heterosexual categories, producing a non-uniform yet distributed pattern. 
Conversely, lesbian, bisexual, and gay categories exhibit very low intensity in Engineering across most models compared to heterosexuals, suggesting a consistent heteronormative bias in technical domains.
Overall, these patterns suggest that differentiation by sexual orientation is driven less by hierarchical status (e.g., Sexual Orientation × Social Class remains low–medium bias in Table \ref{tab:bias-results-colored}) and more by domain-specific associations.
}

\textcolor{black}{
Different model families show distinct occupational patterns. 
In Healthcare, GPT models remain relatively flat, with all sexual-orientation groups low and no strong peaks. In contrast, Mistral models assign a much higher share of healthcare roles. More specifically, lesbian representation in Healthcare rises sharply: 24\% in Ministral-3B and 21\% in Mistral-small, while Mistral-medium partly reverses this trend, lowering lesbian dominance.
On the other hand,  GPT‑5 series show a sharp rise in IT \& Software compared to earlier GPT variants. Across GPT‑3.5 and GPT‑4.x, IT allocations are modest (Gay 9.5\%, Lesbian 3.7\%, Bisexual 8.2\%, Heterosexual 10.2\% on average). GPT‑5 series lift these averages substantially (Gay 22.3\%, Lesbian 14.7\%, Bisexual 26.0\%, Heterosexual 16.7\%), signaling a strong family-level shift toward IT domains, especially for non-heterosexual profiles. However, internal heterogeneity persists: GPT‑5 nano peaks for Gay (24\%) and Bisexual (25\%), GPT‑5 mini pushes Gay to 27\% and Bisexual to 26\%, while GPT‑5 reverses leadership with Lesbian at 22\%, surpassing Gay (16\%). A notable insight is the role of Lesbian profiles: historically underrepresented in IT across GPT models, they surge in GPT‑5  (22\%), signaling a late-stage correction or instability rather than a consistent trend.
These patterns reveal that bias is architecture-dependent and variant-sensitive. GPT‑5 series amplify IT associations broadly across non-heterosexual identities, while Mistral reinforces healthcare stereotypes for lesbian profiles. The variability within GPT‑5 series highlights that bias is dynamic, effective mitigation strategies must therefore be variant-aware, monitor evolving patterns, and address multi-group fairness rather than assuming stability.
}

\subsubsection{Intersectional analysis}

\begin{figure}[h]
  \centering
  \includegraphics[width=\linewidth]{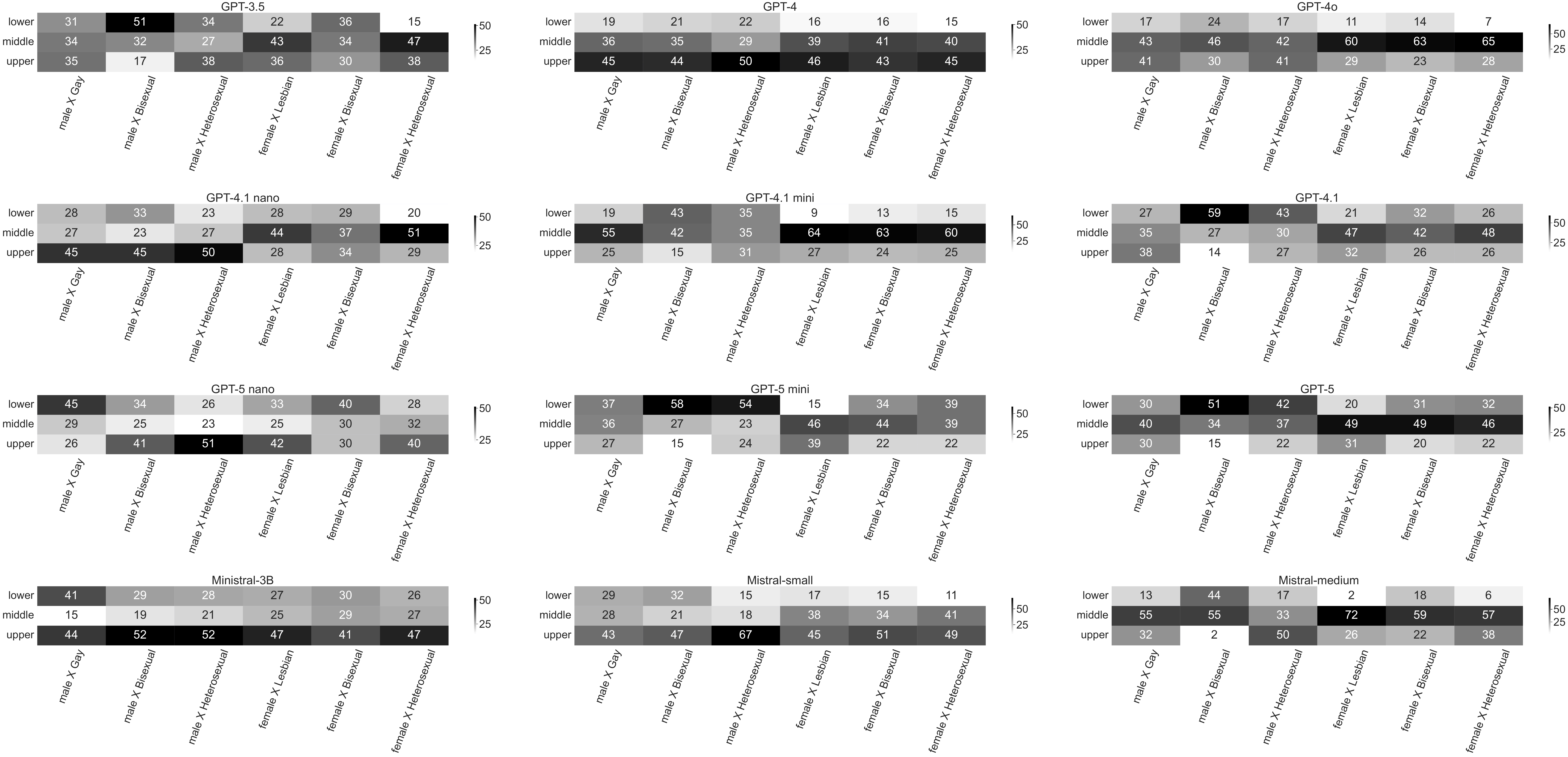}
  \caption{Heatmaps of Gender X Sexual Orientation intersection - Social Class distributions for 12 LLMs. Each subplot corresponds to one model. Each cell is the percentage of that Gender X Sexual Orientation intersection assigned to a given Social Class (columns sum = 100\% per Gender X Sexual Orientation intersection, within model). }
  \Description{Heatmaps showing Gender X Sexual Orientation intersection - Social Class distributions for 12 LLMs. The figure contains 12 subplots, each representing one model. Columns correspond to Gender X Sexual Orientation intersection, and rows represent Social Class categories. Each cell shows the percentage of a given Gender X Sexual Orientation intersection assigned to a social class, with columns summing to 100\% per intersection identity within each model. }
  \label{fig:intersection}
\end{figure}

\textcolor{black}{
A key strength of the proposed PBA lies in its ability to enable intersectional analysis across  identity dimensions. Fairness cannot be meaningfully evaluated through only single-axis perspectives, as overlapping identities often surface disparities that remain  hidden otherwise. Intersectional analysis embodies our broader philosophy that fairness is not a fixed endpoint but a dynamic standard, evolving alongside shifting definitions of “group”. 
}

\textcolor{black}{ To illustrate this capability, we examine the intersection of Gender × Sexual Orientation and its relationship to Social Class. Figure \ref{fig:intersection} displays heatmaps for these intersections across 12 LLMs, with each subplot representing one model. For clarity, we focus on high-frequency identities: male × Gay, male × Bisexual, male × Heterosexual, female × Lesbian, female × Bisexual, and female × Heterosexual, associated to lower, middle, and upper social classes. Each cell shows the percentage of an intersectional identity assigned to a social class, with columns summing to 100\% within each model.
}

\textcolor{black}{
Several global trends can be observed from Figure \ref{fig:intersection}:
\begin{itemize}
    \item Sexual orientation disparities within the same gender are pronounced. For male identities, GPT‑3.5, GPT‑4.1 mini, GPT‑4.1, GPT‑5 mini, GPT‑5, and Mistral‑medium consistently associate bisexual men with lower-class roles at high rates, while gay and heterosexual men show more balanced distribution. 
Similarly, for female identities, bisexual women dominate middle-class representation in most models, whereas heterosexual women and lesbian often cluster toward upper-class associations (especially in GPT-3.5, GPT-4.1, GPT‑5 series and Mistral‑medium). 
\item Gender gaps within each sexual orientation category are also apparent. 
For gay versus lesbian identities, several models (GPT‑4o, GPT‑4.1 nano, GPT‑4.1, and Mistral‑medium) consistently associate lesbian identities with middle-class roles, whereas gay identities lean toward upper-class occupations. Interestingly, this trend reverses in GPT‑5 nano and GPT‑5 mini, highlighting model-specific variability.
For bisexual identities, a persistent bias appears across GPT‑3.5, GPT‑4.1 variants, GPT‑5 series, and Mistral‑medium: female intersections dominate middle-class representation, while male intersections exhibit greater variability across social classes, suggesting uneven treatment within the same orientation.
Finally, for heterosexual identities, most models (GPT‑4, GPT‑4o, GPT‑4.1 series, GPT‑5 nano and Mistral series) have higher upper-class associations for male intersections compared to their female counterparts.
\end{itemize}
These findings underscore that intersectional analysis is essential for complementing single-axis metrics, as it reveals disparities that remain invisible when identities are examined in isolation.  Disparities emerge both within gender and within sexual orientation, amplifying bias in nuanced ways. Larger or newer models do not eliminate these patterns; in some cases, they intensify them, underscoring the need for intersectional auditing in real-world deployments.
}

\subsection{RQ3: How do biases emerge, attenuate, or resurface across successive LLM versions?}

\begin{figure}[h]
  \centering
  \includegraphics[width=0.95\linewidth]{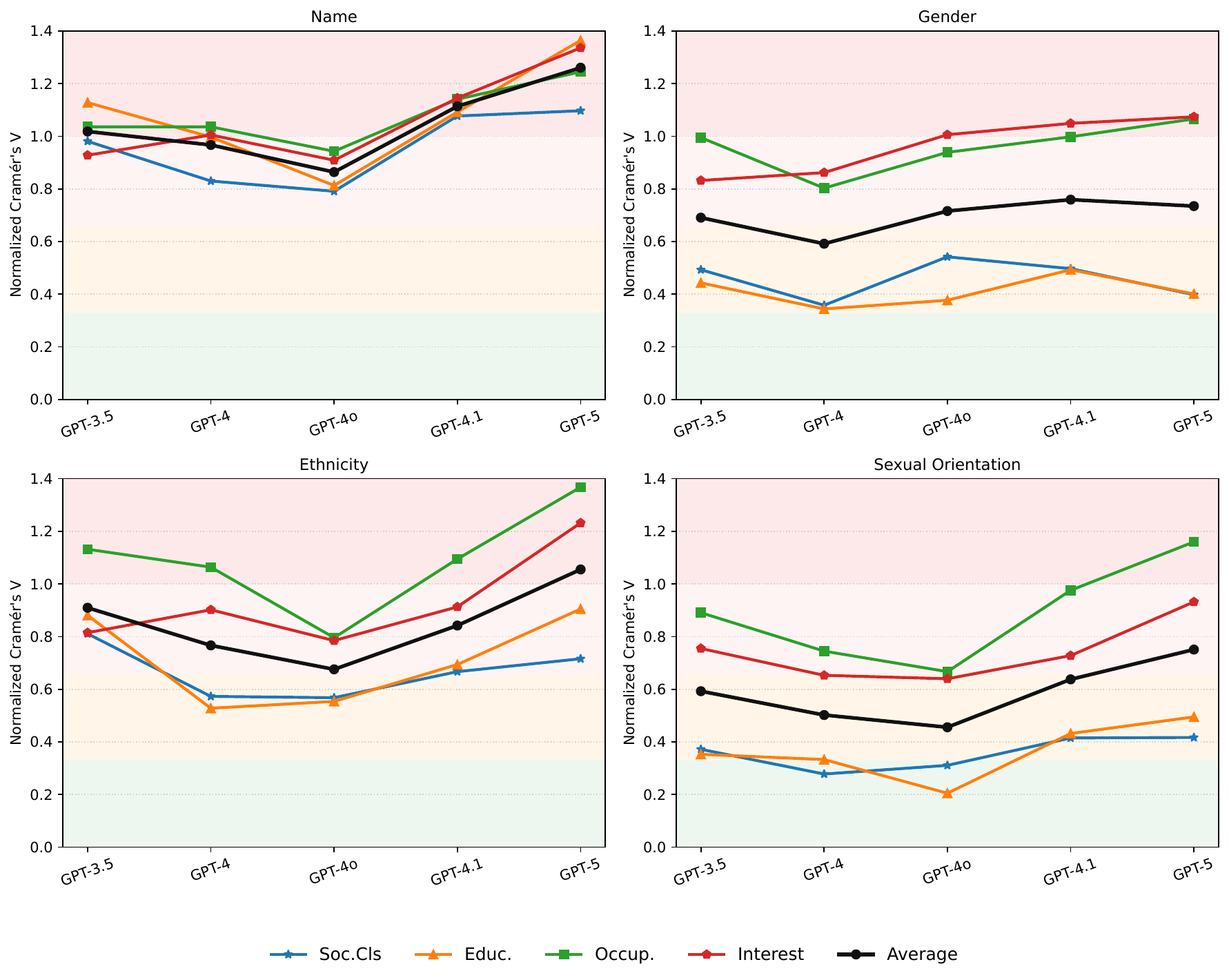}
  \caption{The bias evolution across five model generations (GPT‑3.5 → GPT‑4 → GPT‑4o → GPT‑4.1 → GPT‑5) for four identity axes: Name, Gender, Ethnicity, and Sexual Orientation. Each subplot corresponds to one identity axis and plots normalized Cramér’s V scores for four social dimensions (Social Class, Education, Occupation, and Interest) alongside their average (black line). Shaded bands indicate interpretive thresholds: small (<0.33), medium (0.33–0.66), high (0.66–1.0), and very high (>1.0). }
  \Description{Line charts illustrating bias evolution across five GPT model generations (GPT‑3.5, GPT‑4, GPT‑4o, GPT‑4.1, GPT‑5) for four identity axes: Name, Gender, Ethnicity, and Sexual Orientation. The figure contains four subplots, one per identity axis. Each subplot plots normalized Cramér’s V scores for four social dimensions (Social Class, Education, Occupation, and Interest) along with an overall average shown as a black line. Shaded horizontal bands indicate interpretive thresholds: small (<0.33), medium (0.33–0.66), high (0.66–1.0), and very high (>1.0). The visualization highlights non-linear and sometimes increasing bias trends across model generations, with name and ethnicity biases remaining consistently high and peaking in GPT‑5.
 }
  \label{fig:rq3}
\end{figure}

Bias in LLMs is dynamic rather than fixed. As models evolve through new releases, fine-tuned variants, and fairness interventions, their bias profiles shift in complex ways. Prior work shows that mitigation often follows non-linear trajectories, where biases attenuate in one dimension but resurface in another due to changes in training data, optimization objectives, or alignment strategies \cite{haque2025comprehensive}. Understanding how stereotypes emerge, diminish, or reappear across versions is therefore critical to assess whether fairness gains are systemic or incidental. 

To study how bias envolves across model updates, we focus on five OpenAI models: GPT‑3.5, GPT‑4, GPT‑4o, GPT‑4.1, and GPT‑5, because they represent a clear evolutionary path, offering a natural lens for studying longitudinal bias dynamics. 
These models span major architectural and training shifts.  GPT‑4 introduced multimodal input and achieved human-level performance on benchmarks, far surpassing GPT‑3.5 \cite{openai_gpt4}. GPT‑4o enabled real-time multimodal interaction across text, audio, image, and video \cite{openai_gpt4o}, while GPT‑4.1 improved coding, instruction following, and long-context understanding \cite{openai_gpt41}. The most recent GPT‑5 unifies fast and deep reasoning, reduces hallucinations, and enhances performance across key domains \cite{openai_gpt5}.

Figure \ref{fig:rq3} visualizes how normalized Cramér’s V scores shift across five model generations for four identity axes. Each subplot corresponds to one identity axis and plots normalized Cramér’s V scores for four social dimensions (Social Class, Education, Occupation, and Interest) alongside their average. 
Figure \ref{fig:rq3} reveals that bias trajectories across GPT generations are neither linear nor uniformly improving, challenging the common expectation that newer models are inherently fairer.
Some biases intensify as models gain capabilities. For example, name and ethnicity-based biases remain high throughout and peak in GPT‑5 most of the times, suggesting that the recent improvement abilities on GPT-5 amplify biases rather than eliminate them. 
Among most bias dimensions (e.g., name, ethnicity and sexual orientation), there is a clear bias reduction from GPT-3.5 to GPT-4 to GPT-4o. 
Bias often rebounds in later versions as seen in GPT‑4.1 and GPT‑5. 
These patterns indicate that fairness improvements are not guaranteed by scale, alignment, or new-release alone. For practitioners and users, this underscores the need for longitudinal audits and transparency across releases.

\section{Robustness and Sensitivity Analysis of PBA}
Ensuring the robustness of auditing method is essential for their practical adoption and scientific validity. While the proposed PBA offers a scalable and interpretable approach to bias detection, its reliability under varying conditions must be rigorously evaluated. In this section, we conduct a series of sensitivity analyses to examine how PBA performs when key factors change, including sample size, prompt framing through role-playing, and explicit debiasing instructions. These analyses address two critical questions: (1) Do observed bias patterns remain stable across different data scales and prompt variations;  (2) Can prompt-level interventions meaningfully alter fairness outcomes? 
By systematically examining these factors, we assess the robustness of PBA’s metrics for deploying scalable and trustworthy bias audits in real-world creative AI applications.

\subsection{Impact of sample size}
In the previous analysis, each LLM was prompted to generate 10,000 personas as the basis for bias and fairness evaluation. To examine the robustness of PBA under varying data scales, we conducted a sensitivity analysis across four sample sizes: 5,000, 10,000, 15,000, and 20,000 personas. To ensure sustainability and real-world relevance, we evaluate only the latest models (three from the GPT‑5 family and three from the Mistral series) reflecting current state-of-the-art deployments.
This comparison allows us to assess whether observed bias patterns remain stable as the number of generated personas increases.
To quantify the robustness of PBA across different sample sizes, we compute the following metrics for each bias dimension: 
\begin{itemize}
    \item Intraclass Correlation Coefficient (ICC) \cite{mcgraw1996forming}: \textcolor{black}{ICC is employed to assess the consistency of normalized Cramér’s V scores across conditions.} We report two forms computed across all sizes: ICC(C,1) for quantifying the degree of consistency among measurements and ICC(A,1) for quantifying the degree of absolute agreement \cite{liljequist2019intraclass}. 
    \item Rank stability: Spearman and Kendall rank correlation coefficient are calculated \cite{kendall1945treatment, schober2018correlation}. Spearman’s $\rho$ captures overall monotonic agreement, while Kendall’s $\tau\_b$ assesses pairwise order consistency and explicitly accounts for ties. Used together, they provide complementary views of rank agreement.
    \item Severity Difference (SD):  For ordinal severity levels (e.g., low, medium, high, very high shown in the previous section), we measure the mean absolute difference among different sample sizes.  

\end{itemize}

\begin{table}
\caption{Robustness of PBA across different sample sizes (5k, 10k, 15k, 20k) with multiple metrics. For ICC(C,1),  ICC(A,1),  Spearman,  Kendall, larger values indicate stronger agreement. For Severity Difference, lower values indicate stronger agreement. }
\label{tab:datasize_robustness}
\begin{tabular}{p{3cm} ccccccc}
\toprule
 &  ICC(C,1) ($\uparrow$) & ICC(A,1) ($\uparrow$) & Spearman ($\uparrow$) & Kendall ($\uparrow$) & SD  ($\downarrow$) \\ \hline

\textbf{Name x Social Class} & 0.974 & 0.974 & 0.905 & 0.778 & 0.000 \\
\textbf{Name x Education} & 0.929 & 0.934 & 0.928 & 0.852 & 0.111 \\
\textbf{Name x Occupation} & 0.828 & 0.746 & 0.833 & 0.674 & 0.000 \\
\textbf{Name x Interest} & 0.960 & 0.927 & 0.943 & 0.867 & 0.000 \\
\textbf{Gender x Social Class} & 0.980 & 0.980 & 0.981 & 0.956 & 0.056 \\
\textbf{Gender x Education} & 0.930 & 0.933 & 0.842 & 0.763 & 0.000 \\
\textbf{Gender x Occupation} & 0.993 & 0.992 & 1.000 & 1.000 & 0.000 \\
\textbf{Gender x Interest} & 0.978 & 0.979 & 0.899 & 0.828 & 0.000 \\
\textbf{Ethnicity x Social Class} & 0.975 & 0.974 & 0.943 & 0.867 & 0.111 \\
\textbf{Ethnicity x Education} & 0.983 & 0.978 & 0.947 & 0.898 & 0.000 \\
\textbf{Ethnicity x Occupation} & 0.996 & 0.988 & 1.000 & 1.000 & 0.000 \\
\textbf{Ethnicity x Interest} & 0.986 & 0.986 & 0.981 & 0.956 & 0.056 \\
\textbf{Sexual Orientation x Social Class} & 0.969 & 0.963 & 0.851 & 0.763 & 0.000 \\
\textbf{Sexual Orientation x Education} & 0.874 & 0.853 & 0.714 & 0.689 & 0.000 \\
\textbf{Sexual Orientation x Occupation} & 0.998 & 0.998 & 1.000 & 1.000 & 0.056 \\
\textbf{Sexual Orientation x Interest} & 0.938 & 0.938 & 0.924 & 0.867 & 0.167 \\
\bottomrule
\end{tabular}
\end{table}

Table \ref{tab:datasize_robustness} reports agreement metrics for varying sample sizes. For ICC(C,1), ICC(A,1), Spearman, and Kendall, higher values indicate stronger consistency, whereas for Severity Difference, lower values indicate better agreement. Overall, most bias dimensions exhibit very high stability across sample sizes, reinforcing the robustness of PBA to data scale variations. Notably, Severity Difference remains unchanged for the majority of dimensions (10 out of 16), suggesting that severity assessments are largely insensitive to sample size. Robustness to sampling variation is critical: it ensures that PBA delivers reliable and comparable audits under practical constraints and enables consistent longitudinal bias tracking without requiring fixed or unsustainable large sample sizes.

\subsection{Impact of role playing}
\textcolor{black}{
Role-playing has emerged as a structured prompt-engineering strategy that instructs the model to “act as” a specific persona, such as a teacher or domain expert, to influence reasoning patterns and output characteristics \cite{tseng-etal-2024-two}. Recent work demonstrates its effectiveness for improving alignment and task performance \cite{kong2023roleplay,wang-etal-2024-rolellm,he-etal-2025-crab}. However, its implications for bias and fairness remain underexplored.}  Hence, we propose to study if expert roles such as a UX researcher have an impact on the proposed PBA, which is  essential for designing robust auditing methods.

To systematically examine the effect of role-playing on bias auditing, we compare persona generation under two conditions: a baseline prompt with no roles specified and a role-conditioned prompt where the LLM is instructed to act as an expert UX researcher. 
Beyond the robustness metrics reported in the previous section, we report paired t-test p-values to assess whether role-playing as an expert UX researcher leads to a statistically significant reduction in bias compared to the original prompt (\textcolor{black}{the paired samples are normalized Cramér’s V scores for each bias dimension under two conditions: original prompt versus modified prompt across all models}). A low p-value (e.g., p <0.05) suggests that role-playing produces a significant improvement in bias scores. 

Table \ref{tab:role_robustness} summarizes the results. Agreement metrics suggest that most bias dimensions maintain high stability when prompts are adapted for role-playing, indicating that the underlying bias patterns are largely preserved. The only exception is the Sexual Orientation X Education dimension, which shows reduced agreement. However, its Severity Difference is 0, implying that while the distribution of bias scores may shift slightly, the severity of biased outcomes remains unchanged. 
Moreover, the t-test results reveal no statistically significant differences between the original and role-playing prompts across all bias dimensions. This finding indicates that role-playing as a prompt variation strategy, does not significantly affect PBA outcomes or bias severity.

\begin{table}
\caption{Impact of role-playing on PBA outcomes. For ICC(C,1), ICC(A,1), Spearman, and Kendall, higher values indicate stronger agreement. For Severity Difference (SD), lower values indicate better consistency. For the paired t-test, p-values are reported; values below 0.05 denote a statistically significant difference between original prompt and role-playing prompt.}
\label{tab:role_robustness}
\begin{tabular}{p{3cm} ccccccc}
\toprule
 &  ICC(C,1) ($\uparrow$) & ICC(A,1) ($\uparrow$) & Spearman ($\uparrow$) & Kendall ($\uparrow$) & SD ( ($\downarrow$) & T-test \\ \hline
\textbf{Name x Social Class} & 0.847 & 0.859 & 0.897 & 0.786 & 0.333 & 0.734 \\
\textbf{Name x Education} & 0.648 & 0.622 & 0.600 & 0.467 & 0.167 & 0.876 \\
\textbf{Name x Occupation} & 0.643 & 0.679 & 0.600 & 0.467 & 0.167 & 0.615 \\
\textbf{Name x Interest} & 0.821 & 0.728 & 0.943 & 0.867 & 0.167 & 0.965 \\ \
\textbf{Gender x Social Class} & 0.915 & 0.912 & 0.943 & 0.867 & 0.167 & 0.843 \\
\textbf{Gender x Education} & 0.745 & 0.726 & 0.371 & 0.333 & 0.000 & 0.869 \\
\textbf{Gender x Occupation} & 0.922 & 0.826 & 0.943 & 0.867 & 0.333 & 0.987 \\
\textbf{Gender x Interest} & 0.914 & 0.924 & 0.829 & 0.733 & 0.000 & 0.684 \\
\textbf{Ethnicity x Social Class} & 0.844 & 0.702 & 0.943 & 0.867 & 0.167 & 0.984 \\
\textbf{Ethnicity x Education} & 0.737 & 0.629 & 0.543 & 0.467 & 0.333 & 0.961 \\
\textbf{Ethnicity x Occupation} & 0.891 & 0.879 & 0.886 & 0.733 & 0.167 & 0.879 \\
\textbf{Ethnicity x Interest} & 0.968 & 0.882 & 0.829 & 0.733 & 0.167 & 0.997 \\
\textbf{Sexual Orientation x Social Class} & 0.711 & 0.705 & 0.486 & 0.333 & 0.000 & 0.837 \\
\textbf{Sexual Orientation x Education} & 0.126 & 0.113 & -0.232 & -0.138 & 0.000 & 0.880 \\
\textbf{Sexual Orientation x Occupation} & 0.983 & 0.983 & 0.943 & 0.867 & 0.167 & 0.815 \\
\textbf{Sexual Orientation x Interest} & 0.569 & 0.610 & 0.812 & 0.690 & 0.167 & 0.597 \\
\bottomrule
\end{tabular}
\end{table}

\subsection{Impact of debiasing prompt}
\begin{table}
\caption{Impact of debiase prompt on PBA outcomes. For ICC(C,1), ICC(A,1), Spearman, and Kendall, higher values indicate stronger agreement. For Severity Difference (SD), lower values indicate better consistency. For the paired t-test, p-values are reported; values below 0.05 denote a statistically significant difference between original prompt and debiasing prompt.}
\label{tab:debiase_robustness}
\begin{tabular}{p{3cm} ccccccc}
\toprule
 &  ICC(C,1) ($\uparrow$) & ICC(A,1) ($\uparrow$) & Spearman ($\uparrow$) & Kendall ($\uparrow$) & SD ( ($\downarrow$) & T-test \\ \hline

\textbf{Name x Social Class} & 0.452 & 0.474 & 0.435 & 0.276 & 0.333 & 0.741 \\
\textbf{Name x Education} & 0.112 & 0.112 & 0.486 & 0.333 & 0.167 & 0.823 \\
\textbf{Name x Occupation} & 0.240 & 0.270 & 0.486 & 0.333 & 0.167 & 0.633 \\
\textbf{Name x Interest} & 0.540 & 0.561 & 0.543 & 0.467 & 0.333 & 0.255 \\
\textbf{Gender x Social Class} & 0.433 & 0.442 & 0.371 & 0.333 & 0.167 & 0.791 \\
\textbf{Gender x Education} & 0.450 & 0.352 & 0.543 & 0.467 & 0.167 & 0.949 \\
\textbf{Gender x Occupation} & 0.278 & 0.263 & -0.029 & -0.067 & 0.333 & 0.861 \\
\textbf{Gender x Interest} & 0.233 & 0.254 & 0.371 & 0.333 & 0.500 & 0.288 \\
\textbf{Ethnicity x Social Class} & 0.471 & 0.314 & 0.371 & 0.333 & 0.500 & 0.975 \\
\textbf{Ethnicity x Education} & 0.105 & 0.073 & 0.029 & 0.067 & 0.500 & 0.949 \\
\textbf{Ethnicity x Occupation} & 0.427 & 0.431 & 0.600 & 0.333 & 0.500 & 0.808 \\
\textbf{Ethnicity x Interest} & 0.828 & 0.844 & 0.714 & 0.467 & 0.333 & 0.709 \\
\textbf{Sexual Orientation x Social Class} & 0.103 & 0.087 & 0.314 & 0.200 & 0.000 & 0.902 \\
\textbf{Sexual Orientation x Education} & 0.174 & 0.076 & 0.174 & 0.138 & 0.167 & 0.988 \\
\textbf{Sexual Orientation x Occupation} & 0.662 & 0.606 & 0.771 & 0.600 & 0.333 & 0.918 \\
\textbf{Sexual Orientation x Interest} & 0.593 & 0.611 & 0.580 & 0.276 & 0.500 & 0.759 \\
\textbf{Mean} & 0.320 & 0.319 & -0.200 & -0.200 & 0.167 & 0.822 \\
\bottomrule
\end{tabular}
\end{table}

\textcolor{black}{
Prompt-based debiasing provides a lightweight alternative to retraining for mitigating bias in LLMs by introducing explicit instructions or balanced exemplars within the prompt \cite{echterhoff-etal-2024-cognitive, li2024prompting}. 
Recent work highlights its advantage for rapid deployment but also reports mixed outcomes: while some interventions reduce stereotypical associations \cite{li2024prompting, furniturewala2024thinking}, others lead to superficial shifts or inconsistent effects across models and contexts \cite{yang2025rethinking}. 
These limitations underscore the need for systematic evaluation. In this paper, we test debiasing prompts within the proposed PBA to assess whether such interventions can meaningfully mitigate bias across multiple identity dimensions without retraining.
Inspired by previous studies \cite{cantini2025benchmarking, furniturewala2024thinking, li2024prompting}, we introduced an additional debiasing instruction to the prompt: “Ensure the 20 user profiles represent diversity in gender, age, ethnicity, socioeconomic background, abilities, and geographic regions. Avoid stereotypes and keep descriptions neutral and inclusive”.  
}

We apply the modified debiasing prompt to generate user profiles and use the same metrics as in the previous section on agreements and statistical difference. 
Table \ref{tab:debiase_robustness} summarizes the findings. Overall, most agreement metrics (apart from Kendall) show moderate or higher consistency (>0.40) across most bias dimensions. 
However, these values are substantially lower than those reported in previous sections, indicating that introducing debiasing prompt increases variability in PBA outcomes.
Notably, dimensions such as Ethnicity X Interest and Sexual Orientation X Occupation exhibit relatively high agreement, indicating robustness in these areas. In contrast, dimensions like Ethnicity X Education and Sexual Orientation X Social Class show very low agreement, suggesting that debiasing prompts may influence how bias manifests in these contexts.

Despite these variations, the Severity Difference remains low (mean SD = 0.31), implying that while model rankings may shift, the overall severity of biased outcomes remain stable. Furthermore, t-test results reveal no statistically significant improvements across any bias dimension (all p > 0.05), indicating that the debiasing prompt does not  mitigate bias in LLMs.
These findings highlight the robustness of PBA and suggest that prompt-level interventions alone are insufficient for meaningful bias reduction, underscoring the need for more systematic approaches beyond prompt engineering.

\section{Discussion}
\subsection{Bias trajectories in creative applications}
Our findings reveal that bias in LLM-generated personas is nonlinear, challenging the assumption that newer or larger models are inherently fairer.  Bias trajectories differ not only in magnitude but also in shape across models, even when structural patterns remain stable.  While early frontier models (e.g., GPT‑4, GPT‑4o) exhibit relatively compact bias profiles, later generations such as GPT‑4.1 and GPT‑5 show increased variability and resurgence of certain biases, particularly along name and ethnicity axes. Bias trajectories also differ by family and scale: smaller models sometimes outperform larger ones in fairness. This suggests that fairness is not a monotonic function of scale or alignment but emerges from complex interactions among training data, tuning strategies, and safety layers. For practitioners, this means model upgrades should be treated as fairness risk events, requiring pre- and post-release audits rather than assuming progress.

\subsection{Implications and potential harms}
Persistent gendered occupational stereotypes and non-binary funneling (identified in Section 4.3) pose significant risks in downstream use cases. Male personas cluster in technical and manual roles, while female personas dominate care and creative domains. Non-binary personas exhibit the narrowest occupational spread, often concentrated in Creative \& Design or IT depending on model family. These allocation patterns risk normalizing exclusionary norms, shaping how characters, roles, and narratives are constructed in downstream applications.
The harms extend well beyond representational imbalance. In synthetic data generation use cases, biased outputs can re-enter training pipelines and amplify inequities over time. In high-stakes creative domains such as educational content or recruitment simulations, these biases risk shaping user perceptions and influencing decision-making in subtle yet consequential ways.

\subsection{From measurement to governance}
\textcolor{black}{
PBA’s robustness and interpretability make it suitable for operational integration into deployment pipelines. It benefits three main groups: model developers, model users, and policy makers and auditors.
\begin{itemize}
    \item Model developers, who train and release LLMs, can use PBA to identify bias early and design mitigation strategies before deployment. This includes adjusting training data, adding counterfactual examples, using reinforcement learning or applying fairness constraints. Tracking bias across versions with longitudinal model cards makes fairness a release criterion, not an afterthought.
    \item Model users, such as data scientists, marketing teams, and educators, can use PBA results in two ways. First, they can select models with lower bias for sensitive tasks by comparing audit scores across versions. Second, they can apply mitigation strategies informed by audit findings: adapting prompts to reduce stereotype activation, filtering outputs, or rebalancing persona sets during post-processing. 
    \item Policy makers and auditors gain interpretable metrics and reproducible outputs that support compliance and certification. Normalized bias scores and documented changes across versions provide evidence for regulatory frameworks and help enforce transparency standards.
\end{itemize}
}
To make these benefits practical, we propose three governance practices:
\begin{itemize}
    \item Longitudinal Model Cards: Report normalized Cramér’s V scores for each dimension and track changes across versions. Include differences from the previous release to highlight regressions.
    \item Establish explicit bias thresholds (e.g., normalized bias score < 1.0) that must be met before deployment. For high-risk domains, enforce staged rollouts to mitigate potential harms.
    \item Green Auditing Protocols: PBA’s sample efficiency could potentially reduce compute overhead, but as model ecosystems grow, sustainability must be a first-class concern. Once  personas are generated for bias auditing, they should be shared across teams or organizations. This avoids redundant generation, ensures consistency in bias evaluation, and enables reproducible comparisons without incurring additional compute costs.
\end{itemize}

These practices align with AI governance frameworks emphasizing accountability, transparency, and sustainability. They also position PBA as a practical tool for compliance and continuous monitoring, bridging the gap between research metrics and real-world accountability.

\subsection{ Limitations and future directions}
\textcolor{black}{
While PBA advances fairness auditing, several important limitations remain:
\begin{itemize}
    \item Language and Cultural Scope: Our evaluation is English-only and focuses on models developed in the U.S. and Europe, limiting cultural and linguistic coverage. Future work should include multilingual audits and models from diverse regions to capture broader norms.
    \item  Structured Output Constraints: PBA uses open-ended generation with structured output for comparability and scalability, which may constrain natural language expression. This design ensures reproducibility but focuses on associational bias in persona attributes rather than semantic framing or sentiment. Future work should explore less restrictive formats and incorporate semantic and sentiment analysis.
    \item Prompt sensitivity: although we used a standardized protocol and tested PBA under variations on key factors including sample size,  role-playing prompt and debiase prompt, prompt framing remains an open challenge. A more systematic study of diverse prompt styles is needed to understand how framing interacts with bias detection and to develop guidelines for robust auditing.
    \item Multimodal Bias: PBA currently excludes multimodal outputs, where text-to-image or text-to-video generation introduces other representational risks. Extending PBA to multimodal personas is a key direction for future work.
\end{itemize}
}

\section{Conclusions}

This work introduced the Persona Brainstorm Audit (PBA), a scalable method for auditing bias in open-ended LLM-generated personas using degree-of-freedom-aware normalized Cramér's V, producing bounded severity labels that enable fair comparison across models and dimensions without classifiers or fixed-answer templates. Applying PBA to 12 LLMs across 120,000 personas and 16 bias dimensions, we find that bias trajectories are nonlinear and architecture-dependent: early frontier models (GPT-4o, GPT-4) are among the least biased, while more recent and larger models (GPT-4.1 mini, GPT-5) rank among the most biased. Dimension-level and intersectional analyses reveal that identities appearing fair in isolation can produce compounding disparities in combination, and that stereotype patterns such as occupational clustering by sexual orientation are persistent and model-family-specific. Robustness analyses confirm PBA's stability under varying sample sizes and prompt conditions, though prompt-based debiasing alone proves insufficient for meaningful bias reduction. A current limitation is PBA's focus on personas in English; future work will extend to multilingual settings and investigate causal mechanisms behind nonlinear bias patterns.


\bibliographystyle{ACM-Reference-Format}
\bibliography{sample-base}


\appendix

\textcolor{black}{
\section{Supporting data}
}
\textcolor{black}{
\subsection{Cardinality Comparison Across Models}
To provide transparency and support reproducibility, we include detailed statistics and visualizations of the data processing pipeline and bias analysis results.
}

\begin{table}[h!]
\caption{Comparison of cardinalities among LLMs for different attributes}
\label{tab:categoies}
\centering
\scriptsize
\begin{tabular}{l ccccccc}
\toprule
 & gender & ethnicity & sexual orientation & social class & education level & occupation & top personal interest \\
\midrule
\textbf{GPT-3.5} & 5 & 42 & 15 & 32 & 100 & 297 & 374 \\
\textbf{GPT-4} & 15 & 72 & 16 & 31 & 107 & 519 & 490 \\
\textbf{GPT-4o} & 5 & 57 & 14 & 23 & 85 & 334 & 368 \\
\textbf{GPT-4.1 nano} & 3 & 120 & 19 & 31 & 101 & 491 & 1445 \\
\textbf{GPT-4.1 mini} & 5 & 68 & 11 & 20 & 72 & 344 & 662 \\
\textbf{GPT-4.1} & 16 & 194 & 20 & 17 & 108 & 670 & 700 \\
\textbf{GPT-5 nano} & 7 & 147 & 12 & 31 & 199 & 906 & 2917 \\
\textbf{GPT-5 mini} & 34 & 713 & 23 & 100 & 482 & 1977 & 3885 \\
\textbf{GPT-5} & 23 & 1128 & 11 & 34 & 689 & 1384 & 1501 \\
\textbf{Ministral-3B} & 3 & 50 & 11 & 20 & 19 & 115 & 126 \\
\textbf{Mistral-small} & 9 & 38 & 12 & 16 & 19 & 172 & 307 \\
\textbf{Mistral-medium} & 3 & 18 & 7 & 11 & 13 & 136 & 211 \\ \hline
\textbf{Post-processing} & 3 & 6 & 5 & 3 & 5 & 18 & 15 \\
\bottomrule
\end{tabular}
\end{table}

\textcolor{black}{
Table \ref{tab:categoies} reports the raw cardinalities of key persona attributes (gender, ethnicity, sexual orientation, social class, education level, occupation, and personal interests) for all 12 LLMs prior to normalization, alongside the reduced cardinalities after post-processing. These results highlight the substantial variability in attribute diversity across models. Such variability underscores the necessity of semantic consolidation for meaningful cross-model comparison. Post-processing reduces cardinalities to a minimal, non-redundant set, ensuring statistical robustness and interpretability. Note: For name-based bias evaluation, we analyze the top 50 most frequent names across all models. This choice ensures statistical reliability by avoiding sparsity, supports cross-model comparability by focusing on shared categories, and maintains interpretability while remaining scalable. Using frequent names provides a robust and representative basis for bias detection without introducing noise from rare or model-specific outliers.}

\subsection{Duplicate Profile Analysis}
\textcolor{black}{
Table \ref{tab:dup} summarizes duplicate counts across 10,000 generated profiles per model. Duplication rates vary significantly, from 3 duplicates for GPT-5 to over 2,000 for Mistral-medium. This finding motivated the inclusion of a deduplication step in our pipeline to mitigate skew and ensure fairness in comparative analysis.}

\begin{table}[h!]
\centering
\scriptsize
\caption{Duplicate counts across 12 LLMs for 10 000 persona profiles generated}

\renewcommand{\arraystretch}{1.1}
\setlength{\tabcolsep}{3pt}
\begin{tabular}{l|c|l|c|l|c|l|c}
\hline
\textbf{Model} & \textbf{Dup.} & \textbf{Model} & \textbf{Dup.} & \textbf{Model} & \textbf{Dup.} & \textbf{Model} & \textbf{Dup.} \\
\hline
GPT3.5       & 63   & GPT4.1       & 19        & GPT5         & 3    & Mistral-medium  & 2060  \\
GPT4         & 33   & GPT4.1 mini  & 246           & GPT5mini     & 37   & Mistral-small  & 443 \\
GPT4o        & 11   & GPT4.1 nano  & 81            & GPT5nano     & 6    & Ministral3B  & 776  \\
\hline
\end{tabular}
\label{tab:dup}
\end{table}

\subsection{Gender X Occupation analysis} 
\begin{figure}[h]
  \centering
  \includegraphics[width=\linewidth]{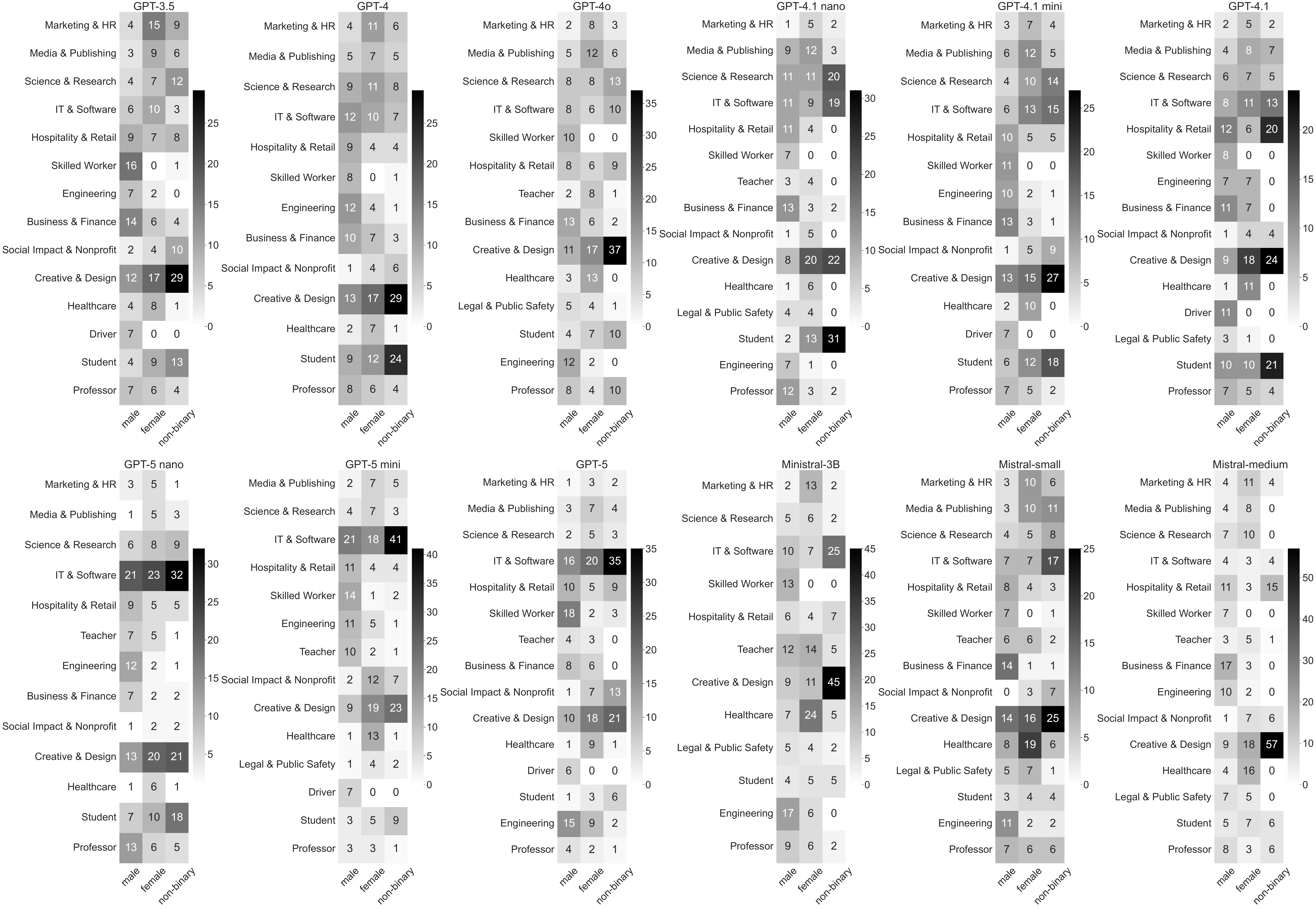}
  \caption{Heatmaps of Gender–Occupation distributions for 12 LLMs . Each subplot corresponds to one model. To facilitate the comparison while focusing on high-mass roles, only top 10 most popular occupations of each gender are selected for each LLM. Each cell is the percentage of that gender assigned to a given occupation (columns sum = 100\% per gender, within model). }
  \Description{Heatmaps showing Gender–Occupation distributions for 12 large language models. The figure contains 12 subplots, each representing one model. For each model, the top 10 most frequent occupations per gender are displayed. Rows correspond to occupations, and columns represent gender categories. Each cell shows the percentage of a given gender assigned to an occupation, with columns summing to 100\% per gender within each model. The visualization highlights differences in occupational allocation patterns across genders and models.}
  \label{fig:gender_occupation}
\end{figure}

To examine how stereotypes manifest across models, we focus on a high-impact dimension: Gender X Occupation. This choice is informed by prior analyses showing that occupational roles are among the strongest correlates of overall bias severity, making them a critical focal point for representational and allocational harms in creative applications.
We visualize in Figure \ref{fig:gender_occupation} the Gender–Occupation distributions for 12 LLMs using heatmaps, with each subplot representing one model. To enable meaningful comparison while emphasizing high-frequency roles, we visualize only the top 10 occupations per gender for each model in Figure \ref{fig:gender_occupation}. Each cell indicates the percentage of a given gender assigned to an occupation (columns sum to 100\% per gender within each model).

Across all twelve models, gendered occupational stereotypes share some common patterns. 
Male personas are repeatedly funneled into technical and manual roles such as Engineering, Skilled Worker, Business \& Finance, and Driver (where present). 
Female personas cluster in creative and care-oriented roles, notably Creative \& Design, Healthcare, Marketing \& HR, and Media \& Publishing. 
Non-binary personas show the narrowest distribution, dominated by Creative \& Design in most models, with a growing shift toward IT \& Software in the GPT‑5 family.
These patterns indicate that gendered differentiation is driven less by hierarchical status (e.g., Gender X Social Class remains low–medium bias level shown in Table \ref{tab:bias-results-colored}) and more by domain-specific associations.

Despite this structural stability, the severity and shape of these biases differ substantially across models. 
Parity between male and female allocations measured by the L1 gap (the sum of the absolute differences between male and female across all occupations) ranges from 54 in GPT‑4 and GPT‑5 nano (most balanced) to 90 in Mistral‑medium (least balanced). 

\textcolor{black}{
Despite structural stability, the severity and distribution of biases vary substantially across models. We quantify gender parity between male and female allocations using the \textit{L1 gap}, defined as the sum of absolute differences between male and female allocations across all occupations:
\begin{equation}
    \text{L1} = \sum_{i=1}^{n} \left| \text{male}_i - \text{female}_i \right|
\end{equation}
where \( n \) is the number of occupations, \text{male}\_i and \text{female}\_i denote the percentage of male and female allocations for occupation \( i \). Higher L1 gap values indicate greater imbalance between genders across occupations. This metric ranges from 54 in GPT‑4 and GPT‑5 nano (most balanced) to 90 in Mistral‑medium (least balanced).}

GPT‑4.1 stands out for achieving near parity in Engineering (7\% male vs. 7\% female) and elevating female and non-binary presence in IT. 
Despite strong overall fairness scores, GPT‑4o retains classic male-coded spikes in Engineering and Business \& Finance as well as female-coded spikes in Creative \& Design and Healthcare. 
The GPT‑5 series introduces a notable shift: non-binary personas move away from creative dominance toward IT, with GPT‑5 mini assigning 41\% of non-binary personas to IT \& Software. 
In contrast, the Mistral family shows increasing creative concentration. In Mistral‑medium, a single category (Creative \& Design) accounts for 57\% of non-binary allocations and the top two categories cover 72\%, indicating extreme funneling.

Family-level trends underscore that bias trajectories are neither linear nor monotonic. 
The GPT‑5 family demonstrates internal heterogeneity: GPT‑5 nano matches GPT‑4 in parity between male and female, while GPT‑5 mini degrades sharply, combining weak parity with the strongest IT funneling for non-binary personas. 
The Mistral family, by contrast, shows consistent creative dominance for non-binary personas, with parity deteriorating as model size increases. These divergences highlight that fairness does not scale predictably with model size or generation.

\subsection{Zoom in analysis} 
\textcolor{black}{
The proposed PBA enables smooth transitions from coarse-grained to fine-grained views by preserving mappings to original terms. After above global analysis of Gender  X  Occupation, which revealed persistent stereotypes on women in creative and care-oriented domains, we zoom into Healthcare to examine the occupation nurse.}

\textcolor{black}{
Table~\ref{tab:nurse} shows gender distribution for nurse across 12 LLMs. Female representation exceeds 90\% in nearly all models, with GPT‑5 reaching the highest male share (12.02\%) and GPT‑4.1 nano assigning the role exclusively to women. Non-binary identities remain marginal, never surpassing 3.38\%.
This example illustrates PBA’s scalability: analysts can move from systemic patterns to occupation-level scrutiny without redesigning the pipeline. Such flexibility is essential for fairness auditing, where biases often surface at different levels of granularity.}

\begin{table}
\caption{Zoom in analysis into Healthcare domain: the gender distribution of nurse occupation across 12 LLMs (each row sums to 100\%). }
\scriptsize
\label{tab:nurse}
\begin{tabular}{lrrr}
\toprule
 & male & female & non-binary \\
\midrule
\textbf{GPT-3.5} & 4.76 & 95.24 & 0.00 \\
\textbf{GPT-4} & 3.07 & 96.93 & 0.00 \\
\textbf{GPT-4o} & 2.06 & 97.94 & 0.00 \\
\textbf{GPT-4.1 nano} & 0.00 & 100.00 & 0.00 \\
\textbf{GPT-4.1 mini} & 1.56 & 98.44 & 0.00 \\
\textbf{GPT-4.1} & 2.27 & 97.41 & 0.32 \\
\textbf{GPT-5 nano} & 5.58 & 93.40 & 1.02 \\
\textbf{GPT-5 mini} & 6.51 & 91.63 & 1.86 \\
\textbf{GPT-5} & 12.02 & 85.79 & 2.19 \\
\textbf{Ministral-3B} & 3.56 & 96.44 & 0.00 \\
\textbf{Mistral-small} & 0.80 & 95.83 & 3.38 \\
\textbf{Mistral-medium} & 9.37 & 90.63 & 0.00 \\
\bottomrule
\end{tabular}
\end{table}

\subsection{Human Validation Study}
\textcolor{black}{
To validate whether higher bias severity indicated by normalized Cramér’s V scores corresponds to greater potential harm, we conducted a comparative human evaluation on the Gender  X  Occupation dimension. This dimension was selected because occupational stereotypes are well-documented and have tangible implications for creative and workplace contexts.
We sampled persona distributions from two LLM variants: System 1 (GPT‑5 mini with a bias score of 1.134, very high bias) and System 2 (GPT‑5 nano with a bias score of 0.779, lower end of high bias). For each model, we aggregated 10,000 generated personas and visualized job distributions by gender across occupational categories. Charts were anonymized and presented as “System 1” and “System 2” to avoid brand or order effects.}

\textcolor{black}{
Participants: 9 evaluators from diverse professional backgrounds (including both DEI experts and non-DEI experts) were recruited to approximate general stakeholder perspectives. Each participant received a concise guideline explaining what patterns to look for (e.g., clustering, coverage, role diversity). They were asked to compare the two systems, assign a Bias Severity Rating, justify their choice, and list the most harmful stereotypes observed.}

Bias Severity Rating: Using the 1–4 scale aligned with our paper:
\begin{itemize}
    \item 	1 = Low Bias (balanced representation, minimal stereotypes).
\item 	2 = Moderate Bias (some clustering, mild stereotypes).
\item	3 = High Bias (clear clustering, multiple stereotypes).
\item	4 = Very High Bias (strong clustering, harmful stereotypes, poor coverage).
\end{itemize}

\begin{table}[ht]
\centering
\caption{Human validation results: bias severity ratings and justifications}
\label{tab:human-validation}
\begin{tabular}{p{2.5cm} p{1.5cm} p{1.5cm} p{8cm}}
\toprule
\textbf{Evaluator} & \textbf{System 1} & \textbf{System 2} & \textbf{Justification} \\
\midrule
Candidate 1 & 3 & 2 & System 1 presents less equal opportunities among genders (especially for engineering, skilled worker, creative jobs); females mostly associated with creative or collaborative roles, males mostly technical. \\ \hline
Candidate 2 & 4 & 3 & Extreme concentration for non-binary personas; reinforcement of traditional stereotypes without meaningful diversity in System 1. \\ \hline
Candidate 3 & 4 & 3 & System 1 has unequal opportunities across genders; may impact career choice if present in chatbot recommending careers. \\ \hline
Candidate 4 & 3 & 2 & High concentration of certain jobs for each gender in System 1, especially for non-binary; lacks diversity. \\ \hline
Candidate 5 & 3 & 2 & Larger gap between distributions in System 1 (e.g., Creative \& Design, Skilled Worker, Driver in male vs female); Social Impact and  Non-profit overrepresented for females. \\ \hline
Candidate 6 & 4 & 3 & System 1 is more biased as: males do technical jobs; females healthcare and NGOs; non-binary mainly IT or creative design (highly clustered). \\ \hline
Candidate 7 & 4 & 3 & Non-binary personas have very low opportunities apart from IT and Creative roles in System 1. \\ \hline
Candidate 8 & 4 & 3 & In System 1, men are mainly associated to technical/manual roles.
while women are more linked to creative and caregiving roles. For non-binary, there is almost exclusively IT \& creative roles. \\ \hline
Candidate 9 & 3 & 2 & In System 1, men are mainly associated to IT related roles and  women are more linked to creative roles, which reinforces existing stereotypes. \\
\bottomrule
\end{tabular}
\end{table}

Table \ref{tab:human-validation} summarizes the results of our exploratory human validation. Across all participants, System 1 was consistently judged as more biased than System 2, with severity ratings clustering at 3–4 for System 1 and 2–3 for System 2. 
Justifications highlight recurring concerns: System 1 exhibits pronounced gender-role segregation, extreme concentration for non-binary personas, and reinforcement of traditional stereotypes (e.g., men in technical roles, women in caregiving and creative domains). 
Several participants noted potential downstream harms, such as influencing career recommendations if these distributions were surfaced in interactive systems. 
While System 2 demonstrates similar tendencies, its distributions are comparatively less polarized, suggesting that lower Cramér’s V scores correspond to reduced perceived harm. These findings provide preliminary evidence that normalized Cramér’s V indeed corresponds to harmful stereotypes and aligns with human judgments of bias severity. 

Limitations: This study serves as an exploratory initiative rather than a definitive validation of bias metrics. Its scope is constrained by a small participant pool, a single bias dimension (Gender  X  Occupation), and a simplified evaluation protocol that does not capture intersectional or cultural nuances. These constraints highlight that validating bias metrics against human perception is a complex research challenge that demands broader sampling, more rigorous methodologies and multi-dimensional analysis. We envision this as a substantial research agenda that warrants a dedicated paper to advance rigorous human-centered validation of bias metrics.

\end{document}